\documentclass[]{iopart}
\usepackage{amssymb}
\usepackage{latexsym}
\usepackage[]{graphicx}
\usepackage{subfigure}
\usepackage[]{hyperref}

\newcommand{\fw}{0.66}

\begin{document}

\title[Spectral Function for Non-Degenerate Coulomb Systems]{Self-consistent Spectral Function for
Non-Degenerate Coulomb Systems and Analytic Scaling Behaviour}
\author{Carsten Fortmann}
\ead{carsten.fortmann@uni-rostock.de}
\address{Institute of Physics, Rostock University, 18051 Rostock, Germany}
\begin{abstract}
  \noindent
  Novel results for the self-consistent single-particle spectral function and self-energy are presented for non-degenerate
  one-component Coulomb systems at various densities and temperatures. 
  The $GW^{(0)}$-method for the dynamical self-energy is used to include many-particle correlations beyond the
  quasi-particle approximation. The self-energy is analysed over a
  broad range of densities and temperatures ($n=10^{17}\,\mathrm{cm^{-3}-10^{27}\,\mathrm{cm^{-3}}}$,
  $T=10^2\,\mathrm{eV}/k_\mathrm{B}-10^4\,\mathrm{eV}/k_\mathrm{B}$). The spectral function 
  shows a systematic behaviour, which is determined by
  collective plasma modes at small wavenumbers and converges towards a quasi-particle resonance at higher wavenumbers. In the low density limit,
  the numerical results
  comply with an analytic scaling law that is presented for the first time. It predicts a power-law 
  behaviour of the imaginary part of the self-energy, $\mathrm{Im}\,\Sigma\propto -n^{1/4}$.
  This resolves a long time problem of the quasi-particle approximation which yields a finite self-energy at vanishing density.
\end{abstract}
\submitto{\JPA}
%
\section{Introduction\label{sec:intro}}
Strongly correlated Coulomb plasmas, found e.g. in planetary interiors \cite{ChabrierSaumon_JPhysA39_4411_2006,Remington:RMP78_755_2006}, 
fusion plasmas \cite{Lindl_PoP_2_3933_1995}, and plasmas excited by lasers or ion beams \cite{Hoffmann_laserparticlebeams23_2004},
are characterized by a high degree
of spatial and temporal correlations, which lead to the emergence of phenomena like collective plasma modes, dynamical screening of the
interparticle interaction potential, and dissolution of bound states. In particular, laser excited plasmas cover a broad range of densities and
plasma temperatures. Values range from typical condensed matter conditions to hot, weakly coupled plasmas.

Theoretical approaches to the physical properties of such systems have to
deal with a great complexity. A particular challenge is the formulation of a coherent theory, which is valid over a wide range of
densities ($n$) and temperatures ($T$), thereby allowing to describe matter in various states, e.g. a
solid-state target, being transferred into a plasma by interaction with high-power lasers and its subsequent relaxation
\cite{ZastrauFortmann_PRL_submitted}. Many-particle perturbation theory \cite{FetterWalecka_1971} presents a general 
approach to many-body systems like condensed matter \cite{Mahan:Book},
partially and fully ionized plasmas \cite{krae}, and nuclear matter, to mention only a few. Also for non-abelian systems, such as the quark-gluon plasma
\cite{0954-3899-20-12-005}, there exist similar approaches to the one described here for Coulomb systems, e.g. the
concept of Schwinger-Dyson equations, see the review article \cite{Hoell:2006AIPC..857...46H}. The thermodynamic properties as well as the
response to external perturbations of these systems in various situations can be studied systematically \cite{zuba1}. 

The central quantity within the many-body theoretical approach is the single-particle spectral function $A(\mathbf{p},\omega)$. 
It represents a physical observable which can be measured via angular resolved photoemission spectroscopy (ARPES)
\cite{1367-2630-7-1-098,0953-8984-10-8-009,0953-8984-10-8-010}. Starting from the spectral function, 
a number of interesting questions related to the physics of many-particle systems can be addressed. 
The equation of state 
\cite{Vorberger_PRE69_046407_2004}, transport cross-sections \cite{KrempSchlangesKraeft:QuantumStatistics} 
(e.g.  electrical conductivity, thermal conductivity, and stopping power \cite{Zwicknagel:PhysRep309_117_1999}), and optical properties \cite{Fortmann:CPP47_2007} 
(emission and absorption of electromagnetic radiation) become accessible. 

In this work, the focus is on the spectral function of plasmas. As an example, a one-component electron
plasma is considered which is charge compensated by a homogeneously distributed background of positively charged ions (jellium model).
The plasma is characterized by the degeneracy parameters $\theta$ and the plasma coupling parameter $\Gamma$ which are defined as
\begin{equation}
  \theta=\frac{2m k_\mathrm{B}T}{\hbar^2 (3\pi^2 n)^{2/3}}\,,\quad\Gamma=\frac{e^2}{4\pi\epsilon_0\,k_\mathrm{B}T}\left(\frac{4\pi n}{3}\right)^{1/3}~.
  \label{eqn:thetadef}
\end{equation}
Here, the electron mass $m$ was introduced, $k_\mathrm{B}$ is the Boltzmann constant. 
In this work, we consider only non-degenerate systems, $\theta\gg 1$, i.e. the thermal energy $k_\mathrm{B}T$ is large compared to the Fermi energy
$E_\mathrm{F}=\hbar^2 (3\pi^2n)^{2/3}/2m$.

The calculation of the spectral function becomes challenging in the regime of
strong coupling, i.e. when the plasma coupling parameter becomes comparable or larger than unity. The coupling parameter measures the ratio of the
Coulomb interaction energy of two particles at a mean distance to  their thermal energy $k_\mathrm{B}T$. At $\Gamma\gtrsim1$, particle
collisions become frequent, involving transfer of both momentum and energy. The interparticle potential is screened due to the presence of nearby
third particles. These correlations significantly modify the plasma observables and
have to be accounted for in the calculation of the spectral function. 
This is accomplished via the
single-particle self-energy $\Sigma(\mathbf{p},\omega)$, 
which is a complex function of both wavevector $\mathbf{p}$ and frequency $\omega$, 
leading to a structured spectral function. 
Though, the main task of many-particle theory, applied to strongly coupled systems, is to calculate the self-energy in a suitable approximation.

The simplest approximation, often found in the literature on Coulomb systems, is the mean-field or Hartree-Fock approximation \cite{krae}. 
One obtains
a frequency independent self-energy which induces a shift in the spectral function's pole, the so-called Hartree-Fock or quasi-particle shift. For dilute plasmas,
this correctly describes the lowering of the chemical potential due to the averaged field of the plasma particles. 
Also, the shift of the ionization energy for bound states is
obtained \cite{EbelingKraeftKremp_1972,SeidelKraeft_PRE52_5387_1995}. However, in dense systems, the mean-field approximation breaks down since
the dynamical screening and collective excitations cannot be accounted for. One has to go beyond the quasi-particle picture.
 
A particularly successful approximation for the self-energy, including these dynamical effects, is the so-called $GW$-approximation
\cite{Aryasetiawan:RepProgPhys61_1998,Mahan:CommCondMatPhys16_1994}. Correlations are accounted for via the dynamically screened
interaction potential $W(\mathbf{q},\omega)$, rather than via the bare Coulomb interaction.
The $GW$ approximation knows a long history of
applications in the field of condensed matter theory. Examples are the calculation of single-particle spectra in the homogeneous electron gas
\cite{Hedin:PhysRev139_1965,BarthHolm:PRB54_1996,Holm:prb57_98}, bandgaps in semiconductors \cite{Northrup:PhysRevLett.59.819}, effective masses
of metal electrons \cite{Godby:PhysRevB.37.10159}, optical and electronic properties of insulators \cite{0953-8984-19-11-116207}, electronic
structure of superconductors \cite{0295-5075-68-6-846}, but also
atomic and molecular systems \cite{0295-5075-76-2-298,0953-4075-28-17-017,1402-4896-1999-T80B-132}. 
In particular, $GW$ self-energy corrections systematically improve band-gap
calculations performed by means of density functional theory \cite{Onida:RevModPhys.74.601,Faleev:PRB74.033101_2006,1367-2630-7-1-126}.

Recently, the $GW$ approximation has been applied also to dense plasmas. Whereas Fehr et al. \cite{FehrKraeft_CPP35_463_1995} performed 
lowest order (one-loop) self-energy corrections to the equation of state,  Wierling et al. \cite{Wierling:CPP38_1998}
carried out pioneering self-consistent calculations of the electron self-energy in the solar-core plasma. An asymmetrically broadened, otherwise
featureless spectral function was obtained.
In this work, the $GW$ self-energy and the corresponding spectral function is investigated for non-degenerate, one-component electron plasmas. 
Only unbound electrons are considered, bound state contributions can be accounted for via T-matrix
calculations, as done in \cite{Schepe-Schmielau:PSSb206_1998}. The self-energy is evaluated for a broad range of densities and temperatures, going
from ideal, weakly coupled plasmas ($\Gamma\ll 1$) to the strong coupling regime $\Gamma\gtrsim 1$. As a novel contribution to the field, an
analytic scaling law for the $GW$ self-energy at low densities is derived which accurately describes the numerical data in this limit.
This expression can be combined with corresponding formulae that are valid in the degenerate case, when $k_\mathrm{B}T\ll E_\mathrm{F}$, to construct a
fit formula for the self-energy which then covers a large portion of the density-temperature plane.

Formerly, analytic expressions for the self-energy have been derived that base on the quasi-particle approximation 
\cite{FennelWilfert_AnnPhysL32_265_1974}. In particular, the completely degenerate electron gas at
$T=0$ was considered, using the plasmon-pole approximation \cite{Lundquist_PhysKondMat6_193_1967}, and also weakly coupled ($\Gamma\ll1$), 
classical plasmas ($E_\mathrm{F}\ll k_\mathrm{B}T$), using the Born approximation for the self-energy \cite{krae}. 
The latter result exhibits several problems: 
The imaginary
part of the quasi-particle self-energy is independent of density and carries a prefactor $\propto 1/\hbar$. Thus, 
there is an unphysical finite damping of single-particle states even in the vacuum and the classical limit $\hbar\to 0$ is 
not defined. On the other hand, from physical arguments, one expects that the self-energy vanishes at zero density and 
that it is a purely classical expression ($\hbar=0$),
when $\theta\gg 1$. This problem has remained unresolved up to now. The real part of the quasi-particle self-energy is well behaved,
i.e. it vanishes at zero density and is purely classical.

The new analytic expression for the self-energy presented in this paper is derived without the quasi-particle approximation, i.e. it is a
non-perturbative result. It is shown that only this non-perturbative treatment leads to an expression that is classical for both the real
and the imaginary part and vanishes exactly in the vacuum limit $n\to 0$. 

The work is organized as follows:
After a brief recapitulation of the single-particle spectral function and the $GW$-method in section~\ref{sec:sf-se},
numerical results for the self-consistent spectral function and self-energy will be discussed in section~\ref{sec:numerical_results}.
Section~\ref{sec:analytic} contains the derivation of the non-perturbative scaling law and comparison to the
numerical results. 
In section~\ref{sec:p0w0}, it will be analyzed why the quasi-particle picture is
incapable to give a physically consistent result for the imaginary part of the self-energy.
Conclusions will be drawn in section~\ref{sec:conclusion}. The appendix contains detailed calculations that are only summarized in the main part of
the paper.


\section{Spectral function and self-energy \label{sec:sf-se}}
The derivation of the $GW$-approximation involves some lengthy manipulations. In this section, only the most relevant formulae
are given, while \ref{app:GW} contains the detailed steps.

Central to the description of electronic properties in a many-body system, which is in thermodynamic equilibrium,
is the thermodynamic electron single particle Green function $G(\mathbf{p},z_\nu)$, defined
at the discrete Matsubara frequencies $z_\nu=(2\nu+1)\pi\mathrm{i}\,k_\mathrm{B}T/\hbar,\quad \nu=0,\pm1,\pm2,\dots$. 
It is related to the single-particle self-energy
$\Sigma(\mathbf{p},z_\nu)$ via Dyson's equation
\begin{eqnarray}
  \nonumber
  G(\mathbf{p},z_\nu)	&=&G^{(0)}(\mathbf{p},z_\nu)+ 
  G^{(0)}(\mathbf{p},z_\nu)\Sigma(\mathbf{p},z_\nu)G(\mathbf{p},z_\nu)\\
  &=&\left[ {G^{(0)}}^{-1}(\mathbf{p},z_\nu)-\Sigma(\mathbf{p},z_\nu) \right]^{-1}~,
  \label{eqn:DysonEqG}
\end{eqnarray}
with the free Green function $G^{(0)}(\mathbf{p},z_\nu)=\left[ \hbar z_\nu-\varepsilon_\mathbf{p}\right]^{-1}$.
Also, the single-particle energy
$\varepsilon_\mathbf{p}=\hbar^2p^2/2m-\mu$ is introduced, $\mu$ is the electron chemical potential.
$G(\mathbf{p},z_\mathrm{\nu})$ contains the thermodynamic properties of a single particle coupled to a thermal bath at a given temperature $T$.
For example, the momentum distribution function
is easily obtained by summation of the Green function over all Matsubara frequencies, 
\begin{equation}
  n(\mathbf{p})=k_\mathrm{B}T\sum_{z_\nu}G(\mathbf{p},z_\nu)~.
  \label{eqn:n_q_sum}
\end{equation}

Instead of the complex Matsubara Green function, it is more convenient to operate on the real valued spectral function
$A(\mathbf{p},\omega)$, defined on the real
frequency axis. It carries the same information as the Green function and is defined via
the spectral representation of the latter,
\begin{equation}
  G(\mathbf{p},z_\nu)=\int_{-\infty}^{\infty}\frac{\mathrm{d}\omega}{2\pi}\frac{A(\mathbf{p},\omega)}{z_\nu-\omega}~.
  \label{eqn:spectral_representation_G}
\end{equation}
Here, $\omega$ is a real valued frequency. 
This relation can be resolved for $A(\mathbf{p},\omega)$, 
\begin{eqnarray}
  \fl
  A(\mathbf{p},\omega)&=&-\lim_{\delta\to 0^+}\,2\,\mathrm{Im}\,G(\mathbf{p},\omega+\mathrm{i}\delta)\\
    \fl
    &=&\lim_{\delta\to0^+}\,\frac{-2\,\mathrm{Im}\,\Sigma(\mathbf{p},\omega+\mathrm{i}\delta)}{\left[
  \hbar\omega-\varepsilon_\mathbf{p}-\mathrm{Re}\,\Sigma(\mathbf{p},\omega)
  \right]^2+
  \left[ \mathrm{Im}\,\Sigma(\mathbf{p},\omega+\mathrm{i}\delta)\right]^{2}}~,
  \label{eqn:dyson_A}
\end{eqnarray}
i.e. the spectral function is obtained after analytic continuation of the Green function from the
Matsubara frequencies to arbitrary complex frequencies as the imaginary part of $G(\mathbf{p},\omega+\mathrm{i}\delta)$, when $\delta$ approaches zero from
positive values. In this way, the sign of the imaginary part of the self-energy is fixed, i.e. $\mathrm{Im}\,\Sigma(\mathbf{p}\omega)<0$ for $\delta>0$.
The real part of the self-energy behaves unambiguous for $\delta=0$.

The spectral function usually exhibits several resonances, including a central peak, 
located at the quasi-particle energy $E_\mathbf{p}$, i.e. the solution of the quasi-particle dispersion
\begin{equation}
  E_\mathbf{p}=\varepsilon_\mathbf{p}+\mathrm{Re}\,\Sigma(\mathbf{p},E_\mathbf{p}/\hbar)~,
  \label{eqn:dispersion}
\end{equation}
accompanied by symmetrically distributed satellites
which are attributed to collective modes in the many-particle system \cite{Lundquist_PhysKondMat6_193_1967}. The width of the resonances in the
frequency domain is commonly
identified with the inverse life-time of these excitations.

Let us first look at the lowest order approximation to the self-energy, the Hartree-Fock term $\Sigma^\mathrm{HF}(\mathbf{p})$. It is given by the
convolution of a non-interacting Green function with
the unscreened Coulomb potential $V(q)=e^2/\epsilon_0q^2\,\Omega_0$ ($\Omega_0$ is a normalization volume),
\begin{eqnarray}
  \Sigma^\mathrm{HF}(\mathbf{p},z_\nu)&=&  -k_\mathrm{B}T\sum_{\omega_\mu,\mathbf{q}} G^{(0)}(\mathbf{p}-\mathbf{q},z_\nu-\omega_\mu)\,V(q)\\
  &=&  \sum_{\mathbf{q}}\left[1-n_\mathrm{F}(\varepsilon_{\mathbf{p}-\mathbf{q}})\right]\,V(q)\equiv \Sigma^\mathrm{HF}(\mathbf{p})~,
  \label{eqn:SigmaHF}
\end{eqnarray}
with the Fermi distribution function $n_\mathrm{F}(\hbar\omega)=\left[ \exp(\hbar\omega/k_\mathrm{B}T)+ 1 \right]^{-1}$.
In the first line, summation takes place over the Bosonic Matsubara frequencies
$\omega_\mu=2\pi\mathrm{i}\mu\,k_\mathrm{B}T/\hbar\,,\quad \mu=0,\pm1,\pm2,\dots$.
The first term $\sum_{\mathbf{q}}\,V(q)$ (Hartree term)
diverges, but it is exactly compensated by the same term from the positive charge background. The second
term (Fock term or exchange term)  gives a finite contribution. Closed expressions can be given in the
case of non-degenerate plasmas \cite{krae,FennelWilfert_AnnPhysL32_265_1974} and completely degenerate Fermi gases \cite{Mahan:Book}.
One finds
$\Sigma^\mathrm{HF}(\mathbf{p})\propto n$ in the high temperature limit ($k_\mathrm{B}T\gg E_\mathrm{F}$) and $\Sigma^\mathrm{HF}(\mathbf{p})\propto n^{1/3}$ in the
quantum degenerate case $k_\mathrm{B}T\ll E_\mathrm{F}$, see \cite{krae} for details.
Thus, the Hartree-Fock self-energy fulfills the physical constraint to vanish at zero density.

The Hartree-Fock term is a real function of momentum, only. The corresponding spectral function is shifted from the free particle dispersion,
\begin{equation}
  A^\mathrm{HF}(\mathbf{p},\omega)=2\pi\,\delta(\varepsilon_\mathbf{p}+\Sigma^\mathrm{HF}(\mathbf{p})-\hbar\omega)~.
  \label{eqn:A_HF}
\end{equation}
No imaginary part of the self-energy appears in this approximation, i.e. the life-time of the Hartree-Fock quasi-particles is infinite.
This is consequence of the mean-field approximation, where no fluctuations of the electric field, i.e. no dynamics of the surrounding plasma
particles are taken into account. 
Recently, also the second order exchange contribution to the self-energy
has been obtained in closed form \cite{1751-8121-40-6-002,Ziesche:psb244_2022_2007}, see also \cite{Ziesche:AnnPhysL_16_45_2006}. However,
this term and all higher order terms, involving only the bare Coulomb potential,
do not lead to a finite particle life-time, only a shift of the dispersion relation is obtained.

To describe the situation in a
dense and strongly correlated system, where the single particle states are spectrally broadened, i.e. they
acquire a finite life-time, one has to go beyond the quasi-particle
approximation, and take into account the screening of the interaction.
The $GW$-approximation, can be regarded as the generalization of the Hartree-Fock theory to dynamically screened interactions. 
It was introduced by Hedin \cite{Hedin:PhysRev139_1965} for the homogeneous electron gas, and is defined as 
\begin{equation}
  \Sigma(\mathbf{p},z_\nu)=
  -k_\mathrm{B}T\sum_{\mathbf{q},\omega_\mu}G(\mathbf{p}-\mathbf{q},z_\nu-\omega_\mu)
  \,W(\mathbf{q},\omega_\mu)~.
  \label{eqn:sigmaGW0_Matsubara}
\end{equation}
$W(\mathbf{q},z)$ is the dynamically screened interaction.
Note that the $GW$ approximation is
a self-consistent ansatz, since the self-energy appears on the l.h.s. as well as in the Green function on the r.h.s. of
(\ref{eqn:sigmaGW0_Matsubara}). Also, the screened interaction $W(\mathbf{q},\omega_\mu)$
is a functional of the Green function via the dielectric function $\epsilon(\mathbf{q},\omega_\mu)$, i.e. the polarization function
$\Pi(\mathbf{q},\omega_\mu)$:
\begin{eqnarray}
  W(\mathbf{q},\omega_\mu)=\frac{V(q)}{\epsilon(\mathbf{q},\omega_\mu)}=\frac{V(q)}{1-V(q)\Pi(\mathbf{q},\omega_\mu)}.
  \label{eqn:W-eps-Pi}
\end{eqnarray}
In $GW$-approximation, $\Pi(\mathbf{q},\omega_\mu)$ is given by the inner product of two Green functions,
$\Pi(\mathbf{q},\omega_\mu)=-k_\mathrm{B}T\sum_{\mathbf{p},z_\nu}G(\mathbf{q}+\mathbf{p},z_\nu+\omega_\mu)\,G(\mathbf{p},z_\nu)$.
The ``double'' self-consistency implied in this ansatz makes the $GW$-approximation is complicated and a numerically demanding problem. 
On the other hand, the full
$GW$-approximation suffers from 
deficiencies due to the neglect
of vertex-corrections \cite{Ward:PhysRev78_1950}, such as violation of the $f$-sum rule \cite{Tamme_PRL83_241_1999}. 
This problem can be avoided by keeping the dynamically screened
interaction on the level of the random phase approximation (RPA) \cite{AristaBrandt:PRA29_1984}, defined by the RPA polarization function,
\begin{eqnarray}
  \Pi_\mathrm{RPA}(\mathbf{q},\omega_\mu)= -k_\mathrm{B}T\sum_{p,z_\nu}
  G^{(0)}(\mathbf{p}-\mathbf{q},z_\nu-\omega_\mu)\,G^{(0)}(\mathbf{p},z_\nu)~,\\
  \Pi_\mathrm{RPA}(\mathbf{q},\omega+\mathrm{i}\delta)=-\sum_\mathbf{k}\frac{n_\mathrm{F}(\varepsilon_{\mathbf{k}+\mathbf{q}/2})-n_\mathrm{F}(\varepsilon_{\mathbf{k}-\mathbf{q}/2})}{\hbar(\omega+\mathrm{i}\delta)+\varepsilon_{\mathbf{k}-\mathbf{q}/2}-\varepsilon_{\mathbf{k}+\mathbf{q}/2}}~.
  \label{eqn:epsPiRPA}
\end{eqnarray}
The use of the RPA polarization function leads to the so-called 
$GW^{(0)}$ approximation for the self-energy. It has been shown to
give more accurate quasi-particle energies \cite{Holm:prb57_98} than the full $GW$-approximation.  
Additionally, it is known that higher order corrections beyond $GW^{(0)}$, such as vertex-corrections and corrections in the polarization function
beyond RPA, partially compensate. Therefore, ignoring them altogether is expected
to give better results than accounting for one or the other
\cite{Mahan:CommCondMatPhys16_1994}. The $f$-sum rule is fulfilled. Further sum rules, e.g. for the moments of the spectral function can be derived
\cite{BarthHolm:PRB54_1996} which are useful to control the numerical treatment of the integral equations to solve.

The inverse dielectric function $\epsilon_\mathrm{RPA}^{-1}(\mathbf{q},\omega)$ describes the propagation of electromagnetic waves in the plasma.
As a main feature, it contains the longitudinal plasma oscillations or plasmons. These resonances show up as peaks in the inverse dielectric
function, located at the roots of the plasmon dispersion
$\mathrm{Re}\,\epsilon_\mathrm{RPA}(\mathbf{q},\omega)=0$. For non-degenerate systems, as considered here, the plasmon dispersion can be
expanded in powers of the wavenumber $q$, and one finds the Gross-Bohm relation \cite{GrossBohm_PhysRev.75.1851}
  $\omega_\mathrm{res}^2(q)=\omega_\mathrm{pl}^2( 1+q^2/\kappa^2)+(\hbar^2 q^2/2m)^2$
for the plasmon resonance frequency $\omega_\mathrm{res}(q)$. Here, the plasma frequency $\omega_\mathrm{pl}$ and the inverse Debye screening length
$\kappa$
\begin{equation}
  \omega_\mathrm{pl}=\left[\frac{n\,e^2}{\epsilon_0m}\right]^{1/2}\,,\quad\kappa=\left[\frac{n\,e^2}{\epsilon_0\,k_\mathrm{B}T}\right]^{1/2}~,
  \label{eqn:wplas_kappa_def}
\end{equation}
have been introduced.
A detailed discussion of the plasmon resonance in dense plasmas can be found in \cite{Thiele_PRE_submitted}. For the present discussion, 
it is important to keep in mind that
the collective plasma excitations are accounted for via the inverse dielectric function in RPA. This is the main advantage of the
$GW^{(0)}$-approximation compared to the mean-field or Hartree-Fock approximation. Depending on the choice of parameters like density
and temperature, these plasmon resonances determine the shape of the self-energy as a function of the frequency and thereby also the spectral
function, where satellites besides the quasi-particle peak indicate coupled electron-plasmon modes, often referred to as plasmarons \cite{Lundquist_PhysKondMat6_193_1967}.

It should be noted at this point that contributions from bound states to the self-energy are not accounted for in this work. The description is
limited to fully ionized plasmas. Bound state contributions can be included using the concept of the T-matrix, see e.g. the work by Schmielau et
al. \cite{Schepe-Schmielau:PSSb206_1998}.

Using the spectral representation (\ref{eqn:spectral_representation_G}) and the screened interaction (\ref{eqn:W-eps-Pi}), the following equation 
for the imaginary part of the self-energy in $GW^{(0)}$-approximation is obtained after summation over the Bosonic Matsubara frequencies
$\omega_\mu$,
\begin{eqnarray}
  \nonumber
  \fl
  \mathrm{Im}\,\Sigma(\mathbf{p},\omega+\mathrm{i}\delta)= 
  \frac{\hbar}{n_\mathrm{F}(\hbar\omega)}\sum_{\mathbf{q}}\int_{-\infty}^{\infty}\frac{\mathrm{d}\omega'}{2\pi}V(q)
  A(\mathbf{p}-\mathbf{q},\omega-\omega')\\
  \times
  \mathrm{Im}\,\epsilon^{-1}_\mathrm{RPA}(\mathbf{q},\omega')\,
  n_\mathrm{B}(\hbar\omega')\,n_\mathrm{F}(\hbar\omega-\hbar\omega')~,
  \label{eqn:sigmacorr_010}
\end{eqnarray}
with the Bose-Einstein distribution function $n_\mathrm{B}(\hbar\omega)=\left[ \exp(\hbar\omega/k_\mathrm{B}T)- 1 \right]^{-1}$.
The real part of the self-energy is obtained by means of Hilbert transform as
\begin{equation}
  \mathrm{Re}\,\Sigma(\mathbf{p},\omega)=\Sigma^\mathrm{HF}_\mathrm{int}(\mathbf{p})+\mathcal{P}\!\int_{-\infty}^{\infty}\!\frac{\mathrm{d}\omega'}{\pi}\,\frac{\mathrm{Im}\,\Sigma(\mathbf{p},\omega')}{\omega-\omega'}~.
  \label{eqn:Kramers-Kronig}
\end{equation}
$\mathcal{P}$ denotes the Cauchy principal value integration,
$\Sigma^\mathrm{HF}_\mathrm{int}(\mathbf{p})$ is the Hartree-Fock self-energy of the interacting system,
\begin{equation}
  \Sigma^\mathrm{HF}_\mathrm{int}(\mathbf{p})=-\hbar\sum_{\mathbf{q}}\,\int_{-\infty}^{\infty}\!\frac{\mathrm{d}\omega}{2\pi}A(\mathbf{p}-\mathbf{q},\omega)n_\mathrm{F}(\hbar\omega) V(q)~.
  \label{eqn:sigmaHF}
\end{equation}
Finally, to close the set of equations, the chemical potential has to be fixed by inversion of the density relation
\begin{equation}
  n(\mu,T)=2\frac{\hbar}{\Omega_0}\sum_{\mathbf{p}}\int_{-\infty}^\infty\!\frac{\mathrm{d}\omega}{2\pi}\,A(\mathbf{p},\omega)\,n_\mathrm{F}(\hbar\omega)~.
  \label{eqn:densityrelation}
\end{equation}
The factor $2$ in front of the r.h.s. stems from the summation over the spin components.
Together with Dyson's equation (\ref{eqn:dyson_A}), (\ref{eqn:sigmacorr_010})-(\ref{eqn:densityrelation}) constitute
a system of non-linear integral equations for the self-energy. 

Besides the normalization of the spectral function
\begin{equation}
  \hbar\int_{-\infty}^{\infty}\frac{\mathrm{d}\omega}{2\pi}\,A(\mathbf{p},\omega)=1~,
  \label{eqn:normalization_A}
\end{equation}
similar sum-rules can be derived also for higher moments of the spectral function \cite{BarthHolm:PRB54_1996}. 
These are independent of the concrete approximation used for the self-energy.
In second order, one obtains an equation relating the first moment of the spectral function to
the interacting Hartree-Fock self-energy (\ref{eqn:sigmaHF}),
\begin{equation}
  \hbar^2\int_{-\infty}^{\infty}\frac{\mathrm{d}\omega}{2\pi}\,\omega\,A(\mathbf{p},\omega)=
  \varepsilon_{\mathbf{p}}+\Sigma^\mathrm{HF}_\mathrm{int}(\mathbf{p})~.
  \label{eqn:A_moment1}
\end{equation}
Similarly, the second moment is related to the Hartree-Fock energy and the frequency integrated imaginary part of the 
self-energy, which is itself a conserved quantity, at least within the $GW^{(0)}$ approximation, see (\ref{eqn:norm_imsigma}) below,
\begin{eqnarray}
 \nonumber
  \hbar^3\!\!\int_{-\infty}^{\infty}\!\!\frac{\mathrm{d}\omega}{2\pi}\,\omega^2\,A(\mathbf{p},\omega)=
  \hbar\!\!\int_{-\infty}^{\infty}\!\!\frac{\mathrm{d}\omega}{\pi}\,\mathrm{Im}\,\Sigma(\mathbf{p},\omega+\mathrm{i}\delta)
  +\left(
  \varepsilon_{\mathbf{p}}+\Sigma^\mathrm{HF}_\mathrm{int}(\mathbf{p}) \right)^2\,.\\
  \label{eqn:A_moment2}
\end{eqnarray}
For the $GW^{(0)}$ self-energy, Holm and von-Barth have found the following identity, relating the integrals over the imaginary part of the
self-energy to the totally integrated response function,
\begin{equation}
  \hbar\int_{\infty}^{\infty}\frac{\mathrm{d}\omega}{\pi}\,\mathrm{Im}\,\Sigma(\mathbf{p},\omega+\mathrm{i}\delta)=
  \hbar\sum_{\mathbf{q}}\int_{-\infty}^{\infty}\frac{\mathrm{d}\omega}{2\pi}V(q)\,\mathrm{Im}\,\epsilon^{-1}_\mathrm{RPA}(\mathbf{q},\omega)~.
  \label{eqn:norm_imsigma}
\end{equation}

In the next section, results for the self-energy will be presented that are obtained via numerical solution of (\ref{eqn:sigmacorr_010}). The
sum rules given above are used 
to check the accuracy of the numerical results.

\section{Numerical results\label{sec:numerical_results}}
The $GW^{(0)}$-approximation is evaluated numerically for various sets of plasma parameters in the following. A typical example of a weakly
coupled ($\Gamma=0.07$), moderately degenerate ($\theta=2.2$) plasma is the plasma at the solar core, with temperatures of 
$T\simeq100\,\mathrm{Ry}/k_\mathrm{B}\simeq 1360\,\mathrm{eV}/k_\mathrm{B}$ and electron densities of $n\simeq 7\times 10^{25}\,\mathrm{cm^{-3}}$
\cite{Bahcall:RMP67.781.1995}. The solar core plasma has
been investigated using the $GW^{(0)}$-method in a number of previous publications, see
\cite{Fortmann:CPP47_2007,Wierling:CPP38_1998,Fortmann:IEEE_2007}. Here, most attention is paid to a systematic analysis of 
the single-particle spectral function and the self-energy over a broad range of densities and temperatures, 
however, sticking to non-degenerate plasmas and neglecting bound states. 
We therefore
start with a plasma temperature that equals the solar core temperature and a density that is 10\% of the solar core electron density.
Later, higher and lower temperatures will be considered as well, i.e. $k_\mathrm{B}T=10\,\mathrm{Ry}$ and $k_\mathrm{B}T=1000\,\mathrm{Ry}$.
Note that $k_\mathrm{B}T$ is always chosen large against typical binding energies of atoms which are usually of the order of several Ry. Thus,
bound states can be neglected.

The numerical solution of equation (\ref{eqn:sigmacorr_010}) is performed by means of an iterative
algorithm, starting from a suitable initialization of the spectral function. Typically, the algorithm converges after 5-10 iterations. 
The threefold integral (\ref{eqn:sigmacorr_010}) is evaluated on a two dimensional grid with roughly 100 nodes in the frequency
coordinate and 10-20 nodes in the momentum coordinate. The angular integral is performed first, followed by the frequency integration and the
integration over the modulus of the wavenumber $q$. The result is checked for consistency in each iteration 
using the sum-rules (\ref{eqn:A_moment1}-\ref{eqn:norm_imsigma}).
Further details concerning the numerical implementation are provided in \cite{Fortmann_inprep}.

\begin{figure}[ht]
  \begin{center}
    \includegraphics[height=.5\textheight,angle=0,clip]{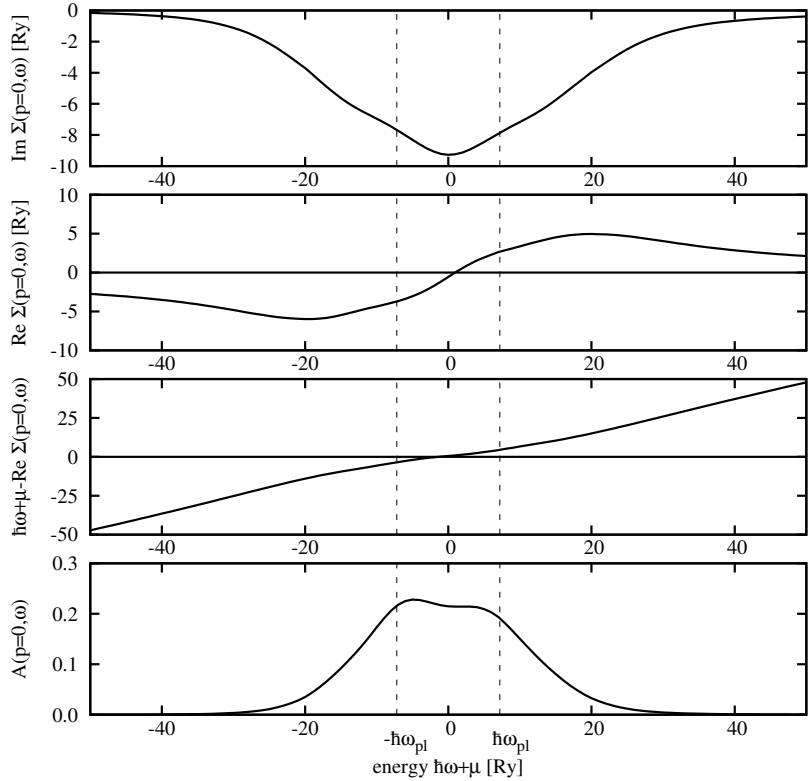}
  \end{center}
  \caption{Self-energy (in units of the Rydberg energy, $1\,\mathrm{Ry}=13.6\,\mathrm{eV}$), dispersion relation (in units of Ry), and spectral
  function (in units of
  1/Ry) for plasma density $n=7\times10^{24}\,\mathrm{cm}^{-3}$ (10\% of the solar core
  density) and temperature
  $T=100\,\mathrm{Ry}/k_\mathrm{B}=1360\,\mathrm{eV}/k_\mathrm{B}$. The spectral function contains two weakly pronounced plasmaron satellites, appearing at
  slightly smaller energies than the plasma frequency (dashed vertical lines). The chemical potential is $\mu=-377
  \,\mathrm{Ry}=-5133\,\mathrm{eV}$. }
  \label{fig:SigmaDispA}
\end{figure}

Figure \ref{fig:SigmaDispA} shows the numerical result for the self-energy (imaginary and real part), dispersion relation
$\hbar\omega+\mu-\hbar^2p^2/2m-\mathrm{Re}\,\Sigma(\mathbf{p},\omega)$, and the spectral function for plasma parameters chosen as $n=7\times10^{24}\,\mathrm{cm^{-3}}$ for
the plasma density, i.e. 10\% of the solar core density,
and $T=100\,\mathrm{Ry}=1360\,\mathrm{eV}/k_\mathrm{B}$ for the plasma temperature. 
The chemical potential is $\mu=-377 \,\mathrm{Ry}=-5133\,\mathrm{eV}$.
The momentum was fixed at $\hbar p=0$. 

The spectral function at the chosen parameters is a broadened resonance with two satellites appearing at about $\hbar\omega+\mu=\pm 5 \,\mathrm{Ry}$
which is slightly below the plasma frequency $\hbar\omega_\mathrm{pl}=7.2\,\mathrm{Ry}$ at the chosen conditions. The plasma frequency is indicated by the
dashed vertical lines. As already mentioned, these satellites are often referred to as plasmarons, i.e. a coupled mode between the single
particle resonance and the collective plasma oscillation \cite{Lundquist_PhysKondMat6_193_1967}.
The imaginary part of the self-energy (top graph) is peaked at the free dispersion $\hbar\omega+\mu=0$ and this peak leads to the small dip in the spectral function
between the satellites, see (\ref{eqn:dyson_A}). The real part of the self-energy (second graph from top) is a rather smooth function, leading to only small variations in the dispersion (3rd graph from top).

\begin{figure}[ht]
  \begin{center}
    \includegraphics[width=\fw\textwidth,angle=0,clip]{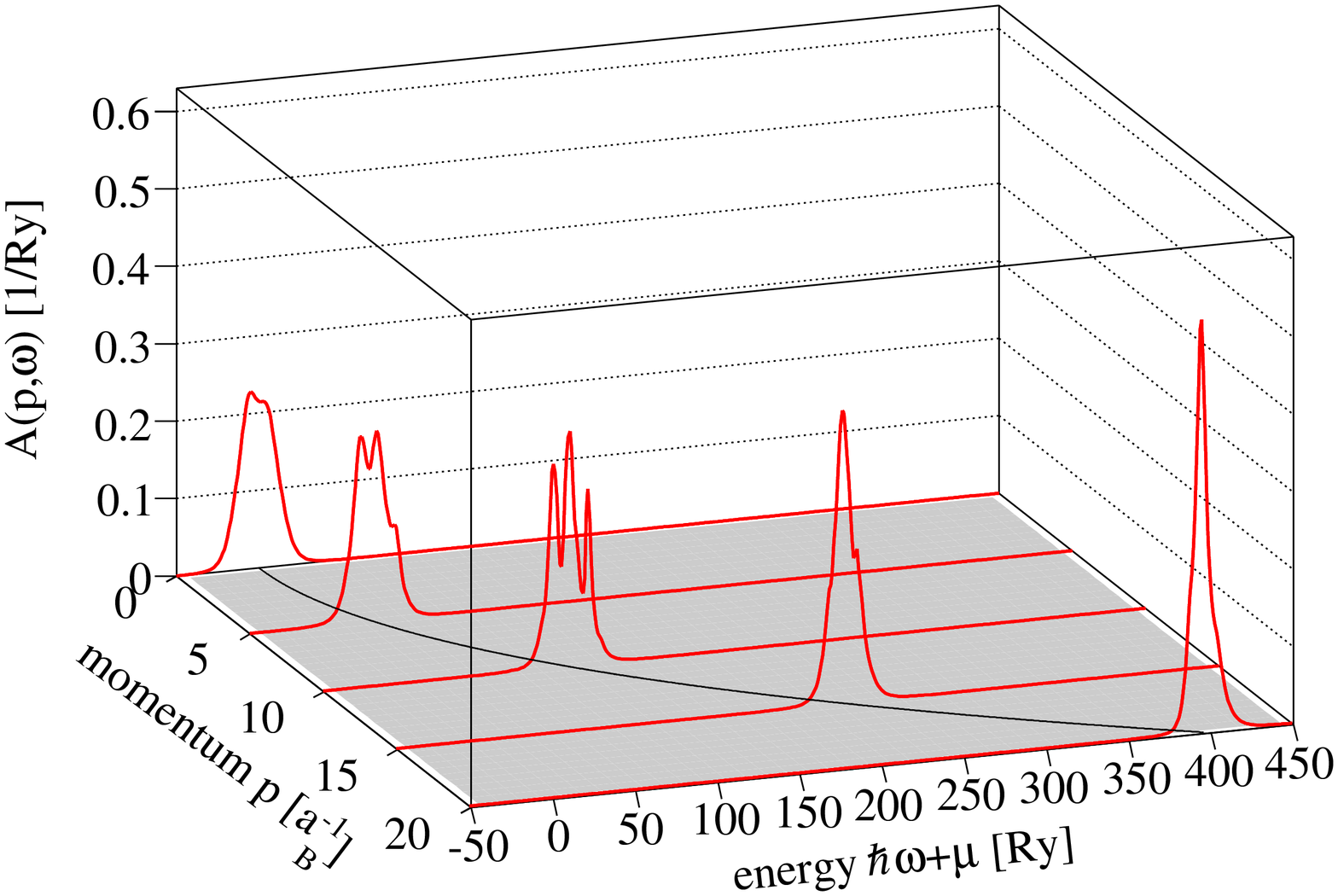}
  \end{center}
  \caption{Spectral function for plasma density $n=7\times 10^{24}\,\mathrm{cm^{-3}}$ and temperature $k_\mathrm{B}T=1360\,\mathrm{eV}$ (solar core
  temperature) as a function of momentum and
  density. The black line on the bottom represents the free dispersion relation $\hbar\omega=\varepsilon_\mathbf{p}=\hbar^2p^2/2m-\mu$. At the
  present parameters the chemical potential is $\mu=-377\,\mathrm{Ry}$. }
  \label{fig:A_3d}
\end{figure}

Next, the dependence of the spectral function on the wavenumber $p$ is analyzed. 
In figure \ref{fig:A_3d}, the spectral function $A(\mathbf{p},\omega)$
is shown for five different wavenumbers, i.e. $p=0,\,5\,a_\mathrm{B}^{-1},\,10\,a_\mathrm{B}^{-1},\,15\,a_\mathrm{B}^{-1},\,\mbox{and
}20\,a_\mathrm{B}^{-1}$,
$a_\mathrm{B}=4\pi\epsilon_0\hbar^2/me^2$ is the Bohr radius.
The density and temperature are the same as before, $n=7\times
10^{24}\,\mathrm{cm^{-3}},\,k_\mathrm{B}T=100\,\mathrm{Ry}$. 
At increased wavenumber ($p\gtrsim 5\,a_{\mathrm{B}}^{-1}$), enhanced complexity of the spectral function is observed.
The plasmaron peaks, which at $\hbar p=0$ appear as small shoulders in the otherwise broad
central resonance, are better defined. The central quasi-particle peak itself becomes 
narrower and the plasmaron peaks separate. At the highest momenta considered ($\hbar p>15\,a_{\mathrm{B}}^{-1}$),
the plasmarons themselves are damped out, and a single, narrow
resonance forms, located near the single particle energy $\hbar\omega=\varepsilon_\mathbf{p}=\hbar^2p^2/2m-\mu$,i.e. the quasi-particle picture 
is restored. Some of these features, especially the plasmaron satellites are already known from literature
\cite{Lundquist_PhysKondMat6_193_1967}.
\begin{figure}[ht]
  \begin{center}
    \includegraphics[width=\fw\textwidth,angle=0,clip]{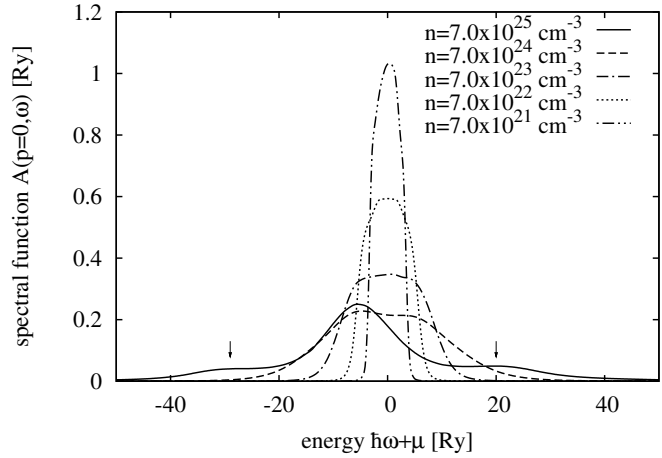}
  \end{center}
  \caption{Spectral functions at $p=0$ for different plasma densities, ranging from the density at the solar core, $n=7\times
  10^{25}\,\mathrm{cm^{-3}}$ (solid curve) to 0.01\% of the solar core (dash-dot-dotted curve). The plasma temperature is
  $T=100\,\mathrm{Ry}/k_\mathrm{B}=1360\,\mathrm{eV}/k_\mathrm{B}$ for all five curves.
  }
  \label{fig:A_comparison_densities}
\end{figure}

Now that the general characteristics of the spectral function have been discussed, the central concern of this paper can be worked out, i.e. the
analysis of the dependence of the self-energy and the spectral function on the plasma parameters density and temperature. 
In figure \ref{fig:A_comparison_densities}, the spectral function at $p=0$ is shown for five different densities between $n=7\times
10^{25}\,\mathrm{cm^{-3}}$ (solar core conditions) and 0.01\% of the solar core density.  The temperature is kept constant at
$T=100 \,\mathrm{Ry}/k_\mathrm{B}=1360\,\mathrm{eV}/k_\mathrm{B}$.
The spectral function drastically changes with varied density. In the case of the highest density considered, 
a narrow quasi-particle peak accompanied by two separate plasmaron satellites (indicated by arrows) is observed. The quasi-particle peak is notably shifted
from the free dispersion $\varepsilon_0=\mu$, due to the real part of the self-energy. Going to lower densities, the plasmaron satellites merge
into the central peak, as can be seen in the case of the spectral function for $n=7\times 10^{24}\,\mathrm{cm}^{-3}$ and also the quasi-particle
shift is reduced. Finally, at the lowest
densities considered, $n=7\times 10^{22}\,\mathrm{cm}^{-3}$ and $7\times 10^{21}\,\mathrm{cm}^{-3}$, a single, narrow quasi-particle resonance
is obtained which is centered around the free dispersion. The width decreases with the density which is the expected behaviour 
in the low density limit.
\begin{figure}[ht]
  \begin{center}
    \includegraphics[width=\fw\textwidth,angle=0,clip]{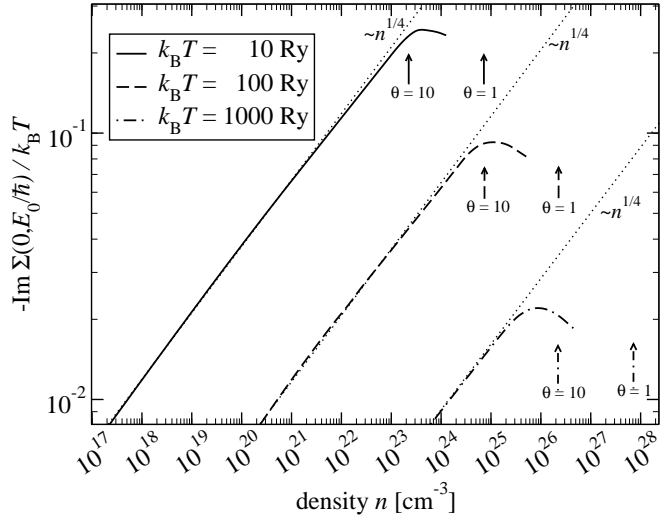}
  \end{center}
  \caption{Effective quasi-particle damping width at $\mathbf{p}=0$, $-\mathrm{Im}\,\Sigma(0,E_0/\hbar)$  normalized to the thermal energy as a function of the plasma
  density. The arrows indicate for each temperature the density at which the degeneracy parameter $\theta$ takes the values 
  $\theta=10$ and $\theta=1$.} 
  \label{fig:effwidth_n_p0}
\end{figure}

In order to study the dependence of the self-energy on density and temperature in more detail, the effective quasi-particle self-energy 
$\Sigma(\mathbf{p},E_\mathbf{p}/\hbar)$ as a function of the density at various temperatures is considered. This quantity gives the shift and width of the central 
peak in the spectral function $A(\mathbf{p},\omega)$, i.e. when $\omega$ is close to the quasi-particle frequency $E_\mathrm{p}/\hbar$, see
(\ref{eqn:dispersion}).
The results for the 
imaginary part of the effective quasi-particle self-energy at $p=0$ as a function of the plasma density are shown in figure \ref{fig:effwidth_n_p0}. Three different temperatures have
been assumed, $T=10, 100,$ and $1000\,\mathrm{Ry}/k_\mathrm{B}$.
Towards low densities, a 
systematic decrease of $-\mathrm{Im}\,\Sigma(0,E_0/\hbar)$ with the density is observed which is
also known from the literature \cite{Wierling:CPP38_1998}. The asymptotes to the low density behaviour, 
shown as thin dotted lines, indicate that $\mathrm{Im}\,\Sigma(\mathbf{p},E_\mathrm{p}/\hbar)$ scales proportional to $-n^{1/4}$. 
This behaviour will be analyzed in more detail
in section \ref{sec:analytic}, where an analytic solution for the $GW$ self-energy is derived that exhibits the same $n^{1/4}$ proportionality.

At higher densities, the
power law behaviour terminates and the self-energy starts to decrease. This can be understood by looking again at figure
\ref{fig:A_comparison_densities}. Here, it was shown that at increased density, the plasmaron satellites separate from the central quasi-particle 
peak, i.e. spectral weight is shifted to the satellites and the central peak narrows.
The calculations have only been
performed for non-degenerate systems, i.e. for densities, where the degeneracy parameter $\theta=k_\mathrm{B}T/E_\mathrm{F}$ is still large
compared to unity. The extension to degenerate systems is straightforward and is covered in another paper \cite{Fortmann_inprep}, 
but will not be treated in this work.

The real part of the self-energy (effective quasi-particle shift), at the densities and
temperatures considered here, was found to follow exactly the Hartree-Fock behaviour,
i.e. $\mathrm{Re}\,\Sigma(0,E_0/\hbar)=-\hbar^2 \kappa^2/2m\propto -n$ \cite{krae}; $\kappa$ is the inverse Debye screening length, see
(\ref{eqn:wplas_kappa_def}). 

\section{Analytic solution for the $GW^{(0)}$ self-energy in Born approximation: Classical limit\label{sec:analytic}}
\subsection{Derivation of the analytic solution}
As discussed,  the spectral function in the low density limit is lacking any plasmaron resonances, only a broadened
quasi-particle peak appears, see figure \ref{fig:A_comparison_densities}. In order to understand this behaviour, the $GW^{(0)}$-equation 
(\ref{eqn:sigmacorr_010}) is reconsidered applying a sequence of approximations as described in the following. In this way, an analytic solution
is found that is valid at low coupling parameters.

It will be shown that the observed scaling is obtained correctly, if the imaginary part of the
self-energy is kept finite also on the r.h.s. of the self-energy integral equation (\ref{eqn:sigmacorr_010}).
It therefore represents a generically non-perturbative result.
Details of the calculations can be found in \ref{app:derivation_analyticSigma}. 

Since collective excitations do not show up in the self-energy and the spectral function at low densities,
it is obvious to neglect these features already in the screened interaction.
Formally, this is achieved by replacing the complete inverse dielectric function by the Born approximation,
\begin{equation}
  \mathrm{Im}\,\epsilon^{-1}(\mathbf{q},\omega)\simeq -\frac{\mathrm{Im}\,\epsilon(\mathbf{q},\omega)}{|\epsilon(\mathbf{q},0)|^2}~.
  \label{eqn:ResponseBorn}
\end{equation}
For the static dielectric function appearing in the denominator, we use the Debye expression $\epsilon_\mathrm{D}(\mathbf{q},0)=
1+\kappa^2/q^2$, with the inverse Debye screening length.
In other words, instead of the interaction via a dynamically screened potential, 
electron-electron collisions via a statically screened potential are considered using the Born approximation. 
Then, (\ref{eqn:sigmacorr_010}) turns into
\begin{eqnarray}
  \fl
  \nonumber
  \mathrm{Im}\,\Sigma(\mathbf{p},\omega+\mathrm{i}\delta)= \\
  \nonumber
  \fl
  \sqrt{\frac{2m\,k_\mathrm{B}T}{\pi^{3}}}\frac{e^2\,\kappa^2}{4\pi\epsilon_0}\!
  \int_{-1}^{1}\!\!\mathrm{d}\!\cos\theta
  \int_{-\infty}^{\infty}\!\!\mathrm{d}\omega'\int_{0}^{\infty}\!\frac{\mathrm{d}q\,q}{\left( q^2+\kappa^2 \right)^2}
  \exp\left( -\frac{m\,\omega'^2}{2q^2\,k_\mathrm{B}T} \right)\exp\left( \frac{\hbar\omega'}{2k_\mathrm{B}T} \right)
  \\
  \fl
  \quad\times
  \frac{\mathrm{Im}\,\Sigma(\mathbf{p}-\mathbf{q},\omega+\mathrm{i}\delta-\omega')}{\left[
  \hbar\omega-\hbar\omega'-\varepsilon_{\mathbf{p}-\mathbf{q}}-\mathrm{Re}\,\Sigma(\mathbf{p}-\mathbf{q},\omega-\omega') \right]^2+\left[
  \mathrm{Im}\,\Sigma(\mathbf{p}-\mathbf{q},\omega+\mathrm{i}\delta-\omega') \right]^2}~.
  \label{eqn:ImSBorn}
  \end{eqnarray}
Note that the dielectric function is taken in the classical limit, i.e. the Fermi-Dirac distribution is replaced by the Maxwell distribution,
leading to the exponentials in the first line of (\ref{eqn:ImSBorn}).

Due to the statically screened Coulomb potential, important contributions to the $q$-integral stem from values
$q\lesssim\kappa$. Therefore, we neglect the shift of momentum in the self-energy on the r.h.s. of equation (\ref{eqn:ImSBorn}), i.e. we
write $\Sigma(\mathbf{p}-\mathbf{q},\omega-\omega')\simeq\Sigma(\mathbf{p},\omega-\omega')$. 
\begin{figure}[ht]
  \begin{center}
    \includegraphics[width=\fw\textwidth,angle=0,clip]{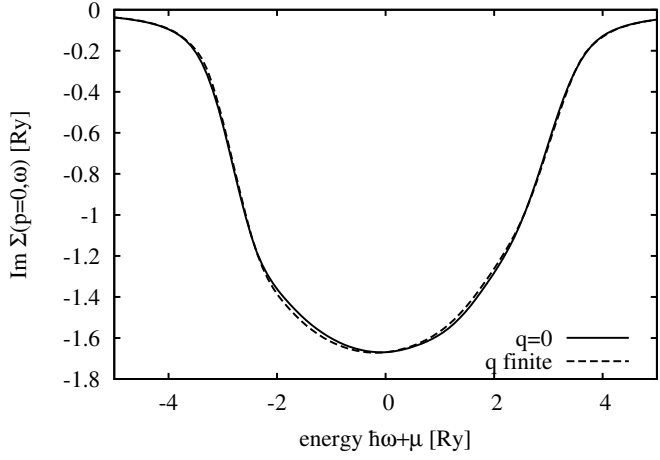}
  \end{center}
  \caption{Imaginary part of the self-energy at momentum $\hbar p=0$ for plasma parameters 
  $n=7\times 10^{21}\,\mathrm{cm^{-3}}$ and $T=100\,\mathrm{Ry}/k_\mathrm{B}$. The self-consistent Born approximation (finite $q$, dashed curve) 
  is compared to the calculation where the momentum shift $\hbar q$ is neglected in the self-energy on the r.h.s. of the self-energy equation
  (solid curve).}
  \label{fig:ImS_q-q0}
\end{figure}
To justify this approximation, 
we show the numerical solution for the imaginary part of the self-energy (\ref{eqn:ImSBorn}) in figure  \ref{fig:ImS_q-q0} (dashed curve). 
The solid curve corresponds to the
solution that is obtained by neglecting the momentum shift in the
argument of the self-energy on the r.h.s. of (\ref{eqn:ImSBorn}). 
As can be seen, this approximation does not modify the result significantly. In fact, the small deviations,
which are only observable around $\hbar\omega+\mu\simeq 0$, are already in the order of the numerical accuracy.

Subsequently, the remaining terms in (\ref{eqn:ImSBorn}) are  expanded in powers of
$q/\kappa$, as described in detail in the appendix. Finally, the threefold integral can be performed and the equation
\begin{eqnarray}
  \fl
 \left[ \mathrm{Im}\,\Sigma(\mathbf{p},\omega+\mathrm{i}\delta) \right]^2+\left[
 \hbar^2p^2/2m-\mu-\hbar\omega+\mathrm{Re}\,\Sigma(\mathbf{p},\omega) \right]^2
 = k_\mathrm{B}T\frac{\kappa\,e^2}{4\pi\epsilon_0}
  \label{eqn:sigma2kappa}
\end{eqnarray}
is obtained.
The l.h.s. is just the denominator of the spectral function, c.f. (\ref{eqn:dyson_A}). Together with the
spectral representation of the Green function (\ref{eqn:spectral_representation_G}), we then find the equation
\begin{eqnarray}
\left[ \hbar z-\hbar^2p^2/2m+\mu-\Sigma(\mathbf{p},z) \right]^{-1}
  &=&
  \frac{4\pi\epsilon_0}{\kappa\, e^2\,k_\mathrm{B}T}\,\Sigma(\mathbf{p},z)~,
\end{eqnarray}
which, in the limit $z=\omega+\mathrm{i}\delta, \delta\to 0^+$ has the solution
\begin{eqnarray}
  \nonumber
  \fl
  \Sigma(\mathbf{p},\omega+\mathrm{i}\delta)=\frac{\hbar\omega-\hbar^2p^2/2m+\mu}{2}\\
  \fl
  \qquad
  -\mathrm{sign}(\hbar\omega-\hbar^2p^2/2m+\mu)
  \left[\left( \frac{\hbar\omega+\mathrm{i}\delta-\hbar^2p^2/2m+\mu}{2} \right)^2-
  \frac{\kappa\, e^2}{4\pi\epsilon_0}k_\mathrm{B}T\right]^{1/2}~.
  \label{eqn:analyticSigma_p-w}
\end{eqnarray}
The signum function,
\begin{equation}
  \mathrm{sign}(\omega)=\left\{\begin{array}{rcl}
    			1&\Leftrightarrow& \omega\geq0\\
			-1&\Leftrightarrow& \omega <0
		      \end{array}\right.~,
\end{equation}
ensures the correct sign of the imaginary part of the self-energy, i.e. $\mathrm{Im}\,\Sigma(\mathbf{p},\omega+\mathrm{i}\delta)<0$ for $\delta>0$.

\subsection{Comparison to the numerical solution}
\begin{figure}[ht]
  \begin{center}
    \subfigure[$p=0$]{\includegraphics[width=\fw\textwidth,angle=0,clip]{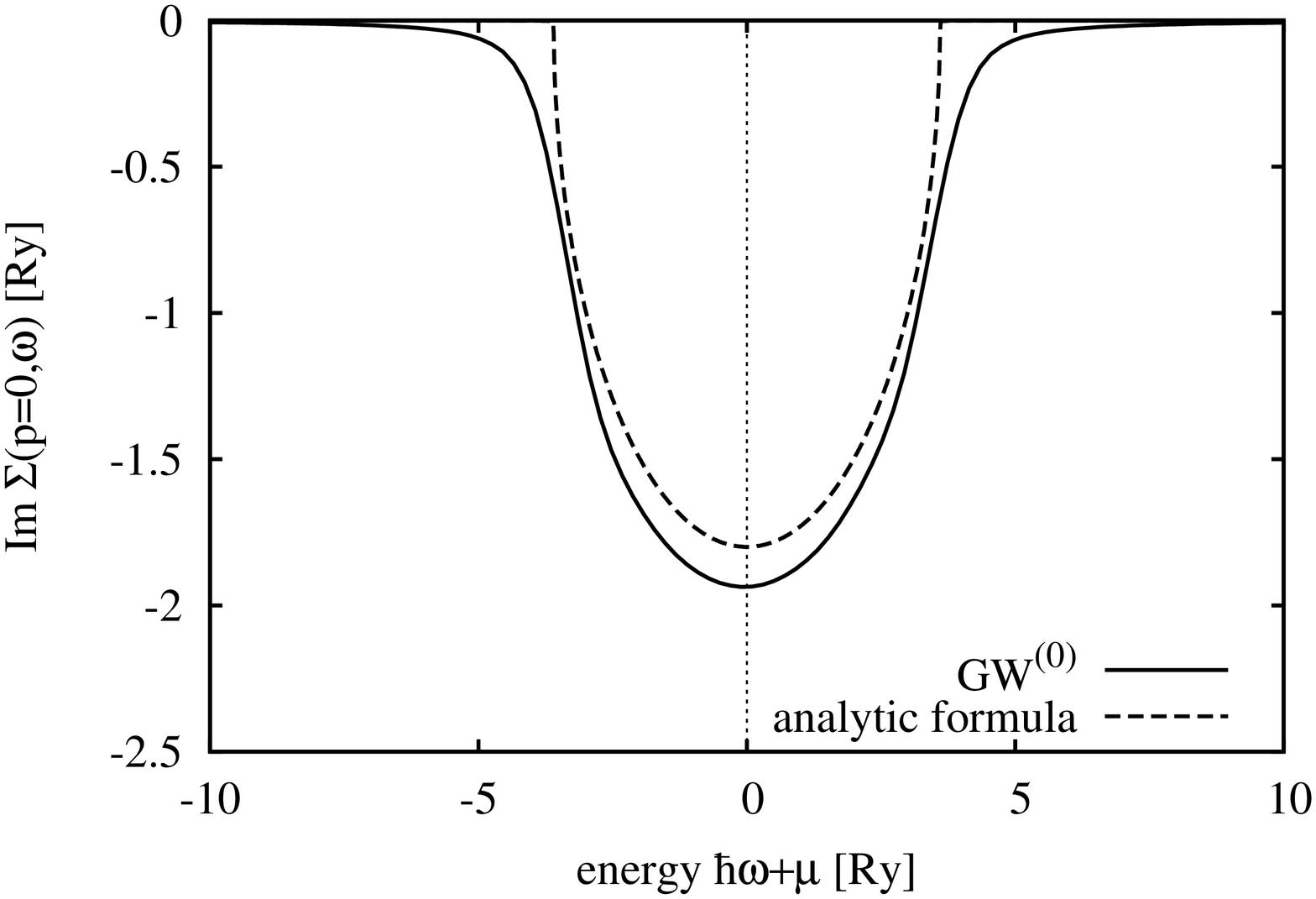}}
    \subfigure[$p=1/a_\mathrm{B}$]{\includegraphics[width=\fw\textwidth,angle=0,clip]{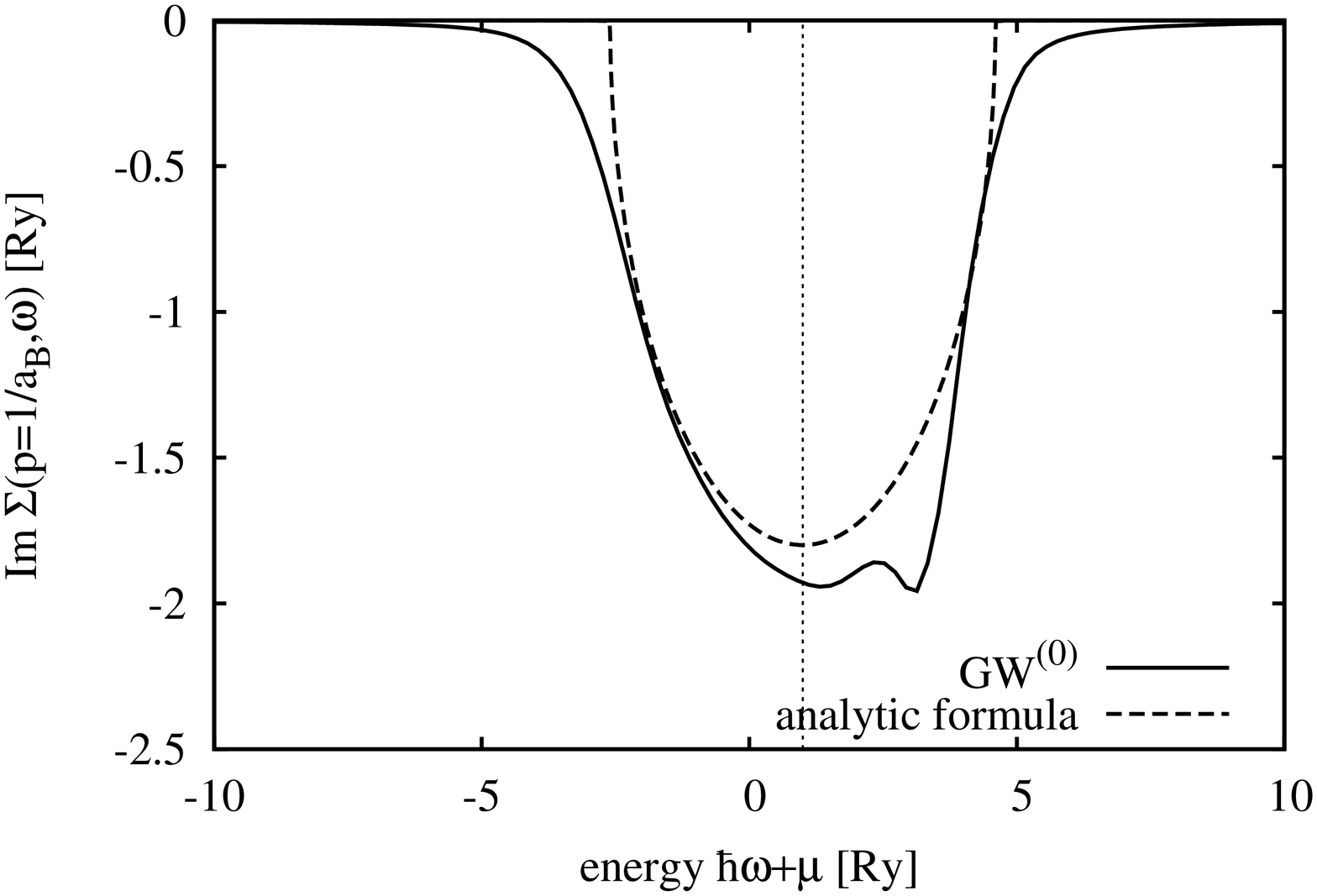}}
  \end{center}
  \caption{Imaginary part of the self-energy for plasma density $n=7\times10^{21}\,\mathrm{cm^{-3}}$ and temperature
  $T=100\,\mathrm{Ry}/k_\mathrm{B}$. Results for $p=0$ (a) and for $p=1/a_\mathrm{B}$ (b) are shown. 
  The self-consistent $GW^{(0)}$-calculation (solid curve) is compared to the analytic formula
  (\ref{eqn:analyticSigma_p-w}) given as dashed curve. The dotted vertical line indicates the quasi-particle dispersion
  $\hbar\omega=E_\mathbf{p}$.}
  \label{fig:ImS_n1e20_T100eV_p0_analytisch-GW}
\end{figure}

The imaginary part of (\ref{eqn:analyticSigma_p-w}) is plotted in figure \ref{fig:ImS_n1e20_T100eV_p0_analytisch-GW}
for $T=100\,\mathrm{Ry}/k_\mathrm{B}$ and $n=7\times10^{21}\,\mathrm{cm^{-3}}$, i.e. for the smallest density considered in figure
(\ref{fig:A_comparison_densities}).  The analytic formula is compared to the full numerical solution for two different wavenumbers,
$p=0$ (a) and $p=1/a_\mathrm{B}$ (b). The dotted vertical line indicates the position of the quasi-particle dispersion
$E_\mathbf{p}$. In the first case, both numerical and analytic calculation agree reasonably well, albeit
the analytic solution lies systematically above the numerical data. However, the overall deviation is smaller than 7\%. In the second case
($p=1/a_\mathrm{B}$), the upshifted plasmon peak, present in the numerical result, is not reproduced by the analytic formula. 
Thus, the analytic formula is
applicable only for small momenta, while at higher momenta, the dynamical features of the interaction become important.

\begin{figure}[ht]
  \begin{center}
    \subfigure[$p=0$]{\includegraphics[width=\fw\textwidth,angle=0,clip]{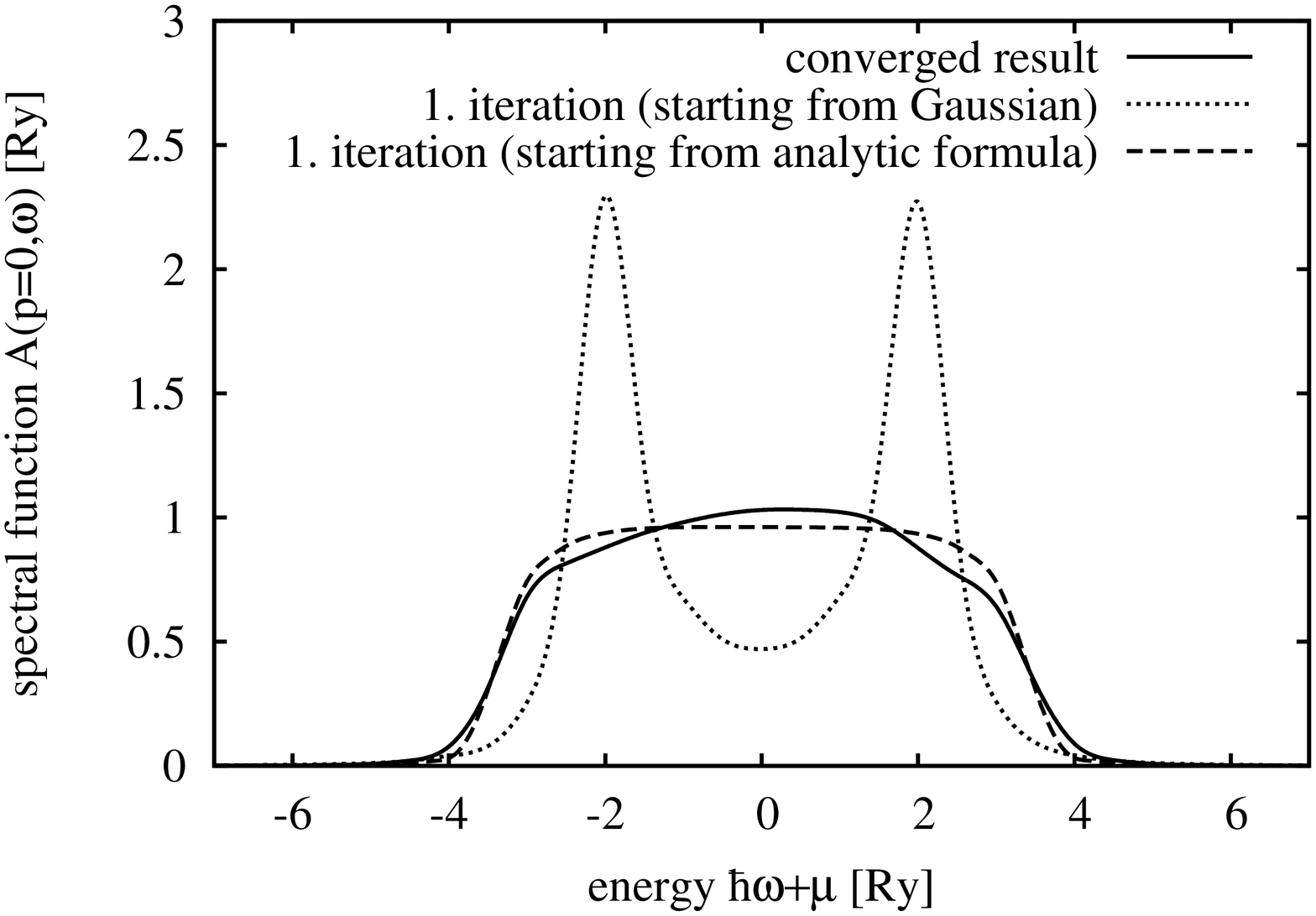}}
    \subfigure[$p=1/a_\mathrm{B}$]{\includegraphics[width=\fw\textwidth,angle=0,clip]{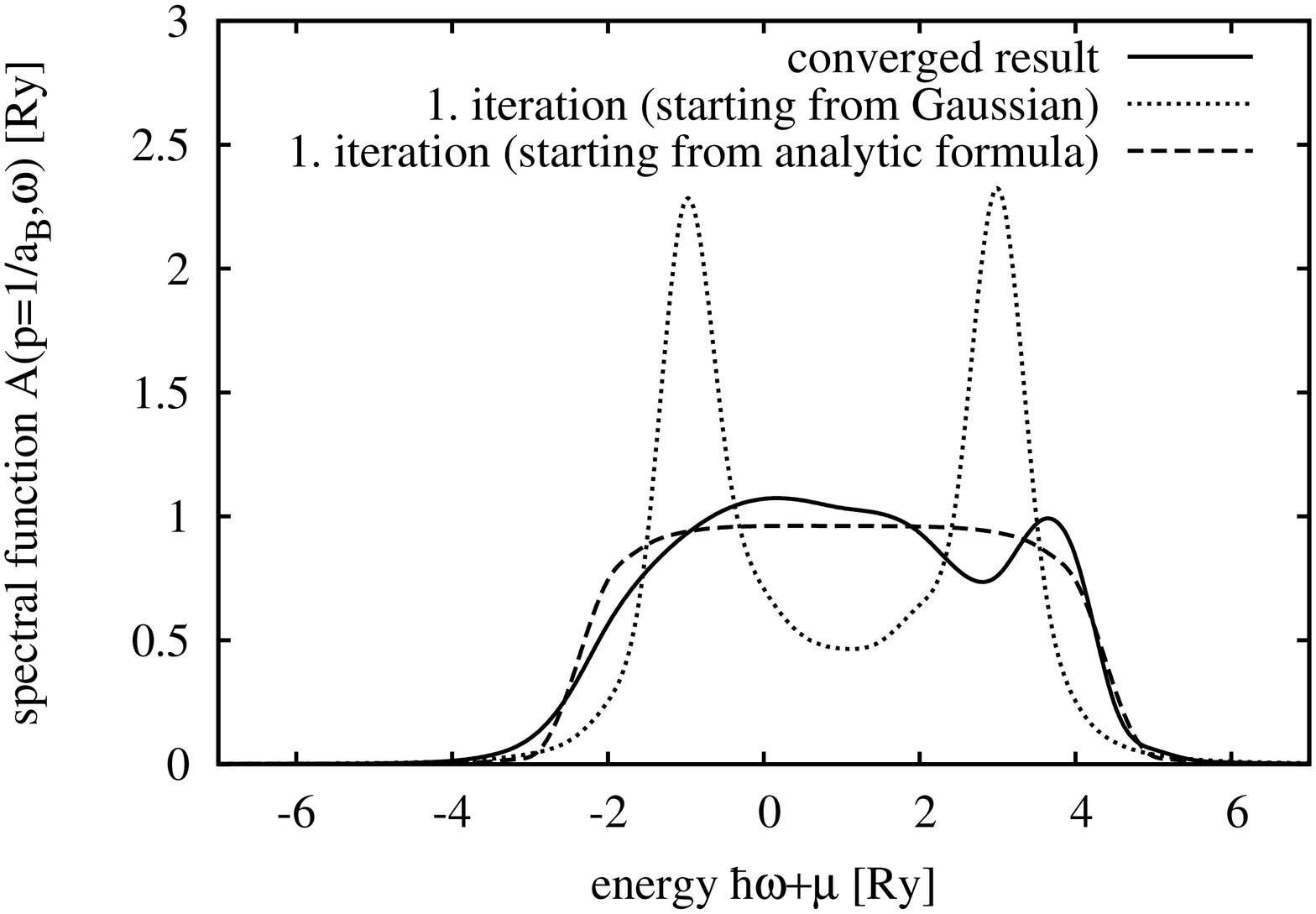}}
  \end{center}
  \caption{Spectral function for plasma density $n=7\times10^{21}\,\mathrm{cm^{-3}}$ and temperature $T=100\,\mathrm{Ry}/k_\mathrm{B}$. Plot (a) shows
  the spectral function at $p=0$, in (b) $p=1/a_\mathrm{B}$ was chosen. The first iteration starting
  from a sharp quasi-particle spectral function (dotted curve) is compared to the first iteration starting from the analytic expression for the
  self-energy (dashed curve).}
  \label{fig:A_QP-analytisch}
\end{figure}

On the other hand,
the analytic formula
is very useful to initialize the numerical algorithm. This is analyzed in 
figure \ref{fig:A_QP-analytisch}. Here, the spectral function, that is obtained in the first iteration of the 
algorithm, was computed in two different ways for the same parameters as above, $n=7\times 10^{21}\,\mathrm{cm^{-3}}$ and
$k_\mathrm{B}T=100\,\mathrm{Ry}$. The dashed curve gives the first iteration starting from
the analytic formula (\ref{eqn:analyticSigma_p-w}) for the self-energy, the dotted curve is the same calculation but starting from a narrow Gaussian
spectral function with a width of $0.3\,\mathrm{Ry}$ (FWHM). 
In plot (a) the wavenumber is $p=0$, while in (b), $p=1/a_\mathrm{B}$ was chosen.
For $p=0$, the analytic ansatz leads to a good resemblance with the converged result (solid curve). 
The converged result is taken here as the 20. iteration starting from the Gaussian ansatz. 
The Gaussian ansatz, iterated once, results in a two-peak structure which is far from the converged spectral function. 
Also at $p=1/a_\mathrm{B}$, starting from the analytic ansatz gives a much better overall correspondence than 
the calculation starting from a Gaussian spectral function, 
although subtle details like the plasmaron peak at $\hbar\omega+\mu\simeq 4\,\mathrm{Ry}$ is not reproduced in the first iteration.

\begin{figure}[ht]
  \begin{center}
    \includegraphics[width=\fw\textwidth,angle=0,clip]{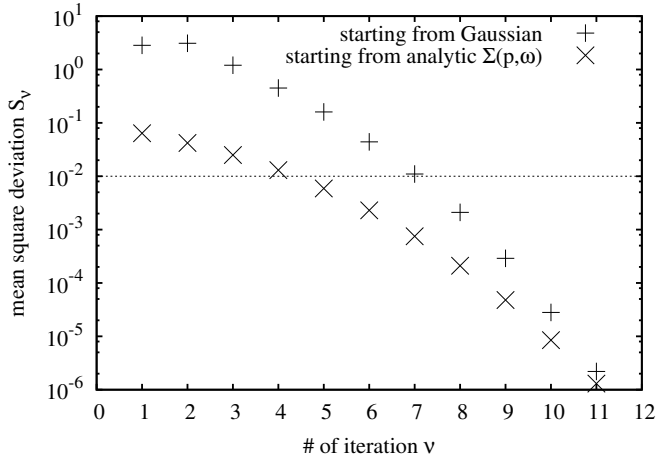}
  \end{center}
  \caption{Mean square deviation of the spectral function at $p=0$ from the converged result as a function of the iteration number $\nu$.
  Red symbols correspond to results obtained when starting with a sharp quasi-particle spectral function, the green symbols result from starting with
  the analytic self-energy and leads to relatively small deviations already in the first iterations. High accuracy ($S<10^{-5}$) is obtained in both schemes
  only after about 10 iterations. Plasma parameters: $n=7\times10^{21}\,\mathrm{cm}^{-3}$, $T=100\,\mathrm{Ry}/k_\mathrm{B}$.}
  \label{fig:convergence}
\end{figure}

In order to perform a quantifiable
comparison between both initializations and their impact on the convergence of the algorithm, we determine the 
mean squared deviation of the spectral function in a given iteration $\nu$
from the converged result $S_\nu=N^{-1}\,\sum_{i=1}^N \left( A^{(\nu)}(0,\omega_i)-A^{(20)}(0,\omega_i) \right)^2$ with $N$
 the number of points on the $\omega$-grid of the
 spectral function, $\omega_i$ the grid points. The result is shown in figure \ref{fig:convergence} for the Gaussian ansatz (marked +)
 and the initialization using the analytic self-energy (marked $\times$).
 During the first four iterations, the mean squared deviation of the second method is by
 two orders of magnitude smaller than if using the Gaussian spectral function. While the mean squared deviation using the analytic self-energy
 becomes smaller than $10^{-2}$ already after  5 iterations, it takes 8 iterations for the Gaussian ansatz to get to this point.
Also, it was found that a Gaussian 
with the width fixed at the imaginary part of the effective quasi-particle self-energy,
does not improve the convergence, since the special form of the self-energy and the spectral function with a broad plateau and 
steep edges cannot be reproduced by such an ansatz and the analytic self-energy given in (\ref{eqn:analyticSigma_p-w}) should be used instead.

\subsection{Analytic solution at the quasi-particle dispersion\label{sec:analyticQP}}
In the following, the analytic solution (\ref{eqn:analyticSigma_p-w}) with the frequency fixed at the quasi-particle dispersion 
$\omega=E_{\mathbf{p}}/\hbar$ shall be considered in more detail. The numerical results for $\Sigma(\mathbf{p},E_\mathbf{p}/\hbar)$ at $\mathbf{p}=0$ have already been discussed in
section \ref{sec:numerical_results}, see figure \ref{fig:effwidth_n_p0}.
Since the only dependence on frequency and wavenumber is given by the trivial
term $\hbar\omega-\hbar^2p^2/2m+\mu=\hbar\omega-\varepsilon_\mathbf{p}$, 
the discussion may be restricted to the case $\mathbf{p}=0$ and
$\hbar\omega=\varepsilon_0=-\mu$. Note that due to (\ref{eqn:analyticSigma_p-w})
$\mathrm{Re}\,\Sigma(\mathbf{p},\varepsilon_\mathbf{p})=0$, therefore
$E_\mathbf{p}=\varepsilon_\mathbf{p}$.
Then, the imaginary part of (\ref{eqn:analyticSigma_p-w}) reads
\begin{eqnarray}
  \mathrm{Im}\,\Sigma(0,-\mu/\hbar)&=-\sqrt{\frac{\kappa\,e^2\,k_\mathrm{B}T}{4\pi\epsilon_0}}
  \label{eqn:ImSanalytisch00}
  =-\left[\left( \frac{e^2}{4\pi\epsilon_0} \right)^{3}4\pi\,n\,k_\mathrm{B}T \right]^{1/4},
\end{eqnarray}
which can also be given in terms of the plasma coupling parameter $\Gamma$,
\begin{equation}
  \mathrm{Im}\,\Sigma(0,-\mu/\hbar)=-\left(3\,\Gamma^3\right)^{1/4}\,k_\mathrm{B}T~.
\end{equation}
This value is the damping width neglecting the influence of
electron-plasmon interaction via the dynamically screened potential. In the following, it is referred to as the non-collective damping width.
The non-collective damping width depends solely on the temperature and the classical coupling parameter $\Gamma$. Therefore, it is a purely
classical result.

We compare this result to the numerical solution for the $GW^{(0)}$ self-energy, presented in section \ref{sec:numerical_results}.
\begin{figure}[ht]
  \begin{center}
    \includegraphics[width=\fw\textwidth,angle=0,clip]{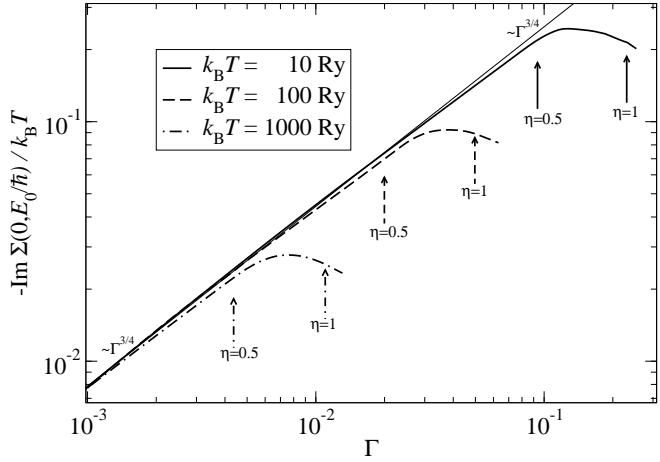}
  \end{center}
  \caption{Effective quasi-particle damping width at $\mathbf{p}=0$, $\mathrm{Im}\,\Sigma(0,E_0/\hbar)$  normalized to the thermal energy as a function of the plasma
  coupling parameter $\Gamma$. The solid black line marks the derived scaling law following (\ref{eqn:ImSanalytisch00}). The vertical lines mark
  for each temperature the onset of plasmon excitation and corresponding decrease of quasi-particle damping, i.e. the parameter
  $\eta=\sqrt{\kappa a_\mathrm{B}}>1$.}
  \label{fig:effwidth_n_p0_scalinglawGamma}
\end{figure}
In figure \ref{fig:effwidth_n_p0_scalinglawGamma}, the effective quasi-particle
self-energy at vanishing momentum $\mathrm{Im}\,\Sigma(0,E_0/\hbar)$, in units of the plasma thermal energy,
is shown as a function of $\Gamma$.
The numerical results perfectly agree with the derived scaling law in the low density limit. 
As long as the density is small, such that the plasma frequency is below the non-collective damping width,
i.e. 
\begin{eqnarray}
  \eta&=\frac{\hbar\omega_\mathrm{pl}}{\sqrt{\kappa\,e^2 k_\mathrm{B}T/4\pi\epsilon_0}}\ll 1
  \label{eqn:eta}
\end{eqnarray}
the damping is mainly non-collective, and a single, broadened resonance appears in the spectral function, c.f.
figure \ref{fig:A_comparison_densities}.

The parameter $\eta$ can also be expressed through the plasma coupling parameter $\Gamma$ and the degeneracy parameter $\theta$, or through the inverse Debye
screening length $\kappa$, 
\begin{eqnarray}
  \eta=
  \left( \frac{2^7}{3^{5/2}\pi^2} \right)^{1/6} \Gamma^{-1/4}\,\theta^{-1/2}
  \simeq 0.9698\,\Gamma^{-1/4}\,\theta^{-1/2}\,,\\
  \eta=\sqrt{\kappa\,a_\mathrm{B}}~.
  \label{eqn:etaGammaTheta}
\end{eqnarray}
Thus, the derived scaling law is only valid for large $\theta$, i.e. classical systems. This was already shown in figure
\ref{fig:effwidth_n_p0}. When $\theta$ approaches 1, quantum effects set in. 
For example, collisions become less probable due to Pauli blocking which leads also to a decrease of the self-energy.

Obviously, the Bohr radius $a_\mathrm{B}$ sets the relevant length-scale that is to be compared to the inverse screening length
$\kappa$ in order to estimate the importance of non-collective damping.
Non-collective damping is the dominant mechanism, as long as the
screening length is large compared to the Bohr radius.
When the screening length becomes smaller than the Bohr radius, i.e. $\eta>1$, which, due to (\ref{eqn:eta}),
is equivalent to having the energy of
plasma oscillations larger than the non-collective damping width,
the plasmaron satellites begin to separate from the broadened quasi-particle peak.
Spectral weight is transferred from the wings of the
central peak into the plasmaron satellites leading to a more defined quasi-particle resonance, i.e. a
decreased damping of the central peak, see figure \ref{fig:A_comparison_densities}.
Concluding, the analytic result is only a good approximation at low densities, when $\eta\ll1$. 
At higher densities, dynamical screening becomes important, leading to satellites in the spectral function.


\section{Deficiencies of the quasi-particle approximation\label{sec:p0w0}}
The non-collective damping width was introduced above in (\ref{eqn:ImSanalytisch00}) as the value of the imaginary part of the self-energy at
vanishing momentum and frequency, $p=0,\,\omega+\mu/\hbar=0$. Of course, the same result is also obtained if this choice of variables was already made at the very
beginning of the calculations leading to (\ref{eqn:analyticSigma_p-w}). However, in the latter case,
the manipulations
can be performed in a different manner. At an intermediate step of the calculation, one can identify the reason
why the quasi-particle damping $\mathrm{Im}\,\Sigma(\mathbf{p},E_\mathbf{p}/\hbar)$ as given in 
\cite{krae,FennelWilfert_AnnPhysL32_265_1974}, behaves unphysical
in the low density and classical limits. 
Detailed calculations are given in \ref{app:p0w0}, while here only the most important steps are summarized.

Setting $\mathbf{p}=0$ and $\hbar\omega+\mu=0$ in (\ref{eqn:sigmacorr_010}), neglecting the momentum shift in the argument of the self-energy on the r.h.s.
and replacing the dynamically screened potential by the statically screened Born approximation as before, we obtain
\begin{equation}
  \fl
  \mathrm{Im}\,\Sigma(0,-\mu/\hbar)=-\frac{e^2\,\kappa^2}{\pi\epsilon_0\,\hbar}\sqrt{\frac{m\,k_\mathrm{B}T}{2\pi}}\,\int_{0}^{\infty}\!\!
  \frac{\mathrm{d}q\,q}{(q^2+\kappa^2)^2}\,\mathrm{Re}\left[ \exp(-z^2)\,\mathrm{erfc}(\mathrm{i}z) \right]\,,\nonumber\\
  \label{eqn:app_scalinglawderiv_030}
\end{equation}
with
\begin{equation}
  z=\frac{\hbar^2q^2+\mathrm{i}\,2m\,\mathrm{Im}\,\Sigma(0,-\mu/\hbar)}{2\hbar q\sqrt{2m\,k_\mathrm{B}T}}~.
\end{equation}

Most contributions to the integral stem from small values of the
wavevector, $q\lesssim \kappa$. Therefore, we may neglect the
real part of $z$ and write
$z=i\,\sqrt{2m}\,\mathrm{Im}\,\Sigma(0,-\mu/\hbar)/2\hbar q\sqrt{k_\mathrm{B}T}$. Using the expansion 
\begin{equation}
  \lim_{x\to+\infty}\,\exp(x^2)\,\mathrm{erfc}(x)=\frac{1}{\sqrt{\pi}\,x}-\frac{1}{2\sqrt{\pi}x^3}+\mathcal{O}(x^{-5})
  \label{eqn:expansion_experfc}
\end{equation}
in lowest order only, the $q$-integral can be performed, resulting in
\begin{equation}
  \mathrm{Im}\,\Sigma(0,-\mu/\hbar)=-\sqrt{\kappa e^2 k_\mathrm{B}T/4\pi\epsilon_0}~,
  \label{eqn:ImSQP}
\end{equation}
which coincides with (\ref{eqn:analyticSigma_p-w}) at $p=0$ and $\hbar\omega=-\mu$.

From  (\ref{eqn:app_scalinglawderiv_030}), one can also derive the quasi-particle approximation for the imaginary part of the self-energy:
If the imaginary part of $z$, i.e. the self-energy, is neglected on the r.h.s. (this is just the quasi-particle approximation),
and furthermore the limit $q\to 0$ is considered, the expression
\begin{eqnarray}
  \nonumber
  \mathrm{Im}\,\Sigma^\mathrm{QP}(0,-\mu/\hbar)&= -\frac{e^2\,\kappa^2}{\pi\epsilon_0\,\hbar}\sqrt{\frac{m\,k_\mathrm{B}T}{2\pi}}\,\int_0^\infty
  \frac{\mathrm{d}q\, q}{(q^2+\kappa^2)^2}\\
  &= -\frac{e^2}{4\pi\epsilon_0\hbar}\sqrt{\frac{2m\,k_\mathrm{B}T}{\pi}}~,
  \label{eqn:QPLimitImS00}
\end{eqnarray}
is obtained.

This coincides with the formula for the imaginary part of the
quasi-particle self-energy as given in \cite[p. 114, equation (4.164)]{krae}. There, the spectral function on the 
r.h.s. of the integral equation for $\Sigma(\mathbf{p},\omega)$ is replaced by an
on-shell delta distribution (free particle spectral function), i.e. the self-energy is set to 0 on the r.h.s.. The resulting integral is evaluated
at the free particle dispersion
$\hbar\omega=\varepsilon_\mathbf{p}$. 

As result, one obtains the expression 
\begin{equation}
  \mathrm{Im}\,\Sigma(\mathbf{p},\varepsilon_\mathbf{p}/\hbar)=-\frac{e^2}{4\pi\epsilon_0\,\hbar}\sqrt{\frac{2m\,k_\mathrm{B}T}{\pi}}\,{}_1F_1(1,3/2;-\varepsilon_\mathbf{p}/2\,k_\mathrm{B}T)~,
  \label{eqn:QPA_krae}
\end{equation}
with ${}_1F_1(\alpha,\beta;z)$ being the confluent hypergeometric function \cite{abra}.
Note that in the given reference, instead of the imaginary part
of the self-energy, the quasi-particle damping
$\Gamma(\mathbf{p},\varepsilon_\mathbf{p}/\hbar)=-2\mathrm{Im}\,\Sigma(\mathbf{p},\varepsilon_\mathbf{p}/\hbar)$ is given. Also, the original
formula differs from (\ref{eqn:QPA_krae}) by a factor of 1/4. However, the formula given here was approved through private communication by W.-D. Kraeft.

Obviously, (\ref{eqn:QPA_krae}) is independent of density. The neglect of $\mathrm{Im}\,\Sigma(0,-\mu/\hbar)$ in the complex variable
$z$ leads to a different analytical structure of
the equation. Therefore,
the quasi-particle approximation has no chance to ever obtain the correct behaviour at low densities. Low densities, and
therefore  small inverse screening lengths $\kappa$ shift the supporter of the $q$-integral to small $q$, where contributions from
$\mathrm{Im}\,z$ are important, 
whereas the real part of $z$ vanishes at $q=0$ and leads to a result which is independent of $\kappa$.

In the same way, one can understand why the quasi-particle limit diverges when considering the \textit{classical} limit $\hbar\to 0$.
The imaginary part of $z$ has $\hbar$ in the denominator which after the integration cancels the $\hbar$ in the prefactor in equation
(\ref{eqn:app_scalinglawderiv_030}). No cancellation takes place, if the imaginary part is neglected, i.e. in the quasi-particle approximation. 
This leads to the divergence of the final result.

\section{Conclusion\label{sec:conclusion}}
In this work, the single-particle self-energy of the one-component electron plasma was investigated. The spectral function was calculated
self-consistently using the $GW^{(0)}$-approximation which allows the systematic treatment of dynamical correlations in the plasma.
The spectral function contains at small momenta a broadened quasi-particle peak and two plasmaron satellites 
which, at low densities, merge into the central quasi-particle resonance. At increased momenta, for a
given density and temperature, the spectral function converges to a single, sharp quasi-particle resonance.
Special attention was paid to a systematic investigation of the self-energy and the spectral
at different densities and temperatures. Here, only non-degenerate plasmas were considered, i.e. the temperature is large compared to
the Fermi temperature. Also, bound states were neglected.

It was found that at low
densities, the imaginary part of the on-shell self-energy, i.e. the inverse
single particle lifetime, follows a universal scaling law $\mathrm{Im}\,\Sigma(\mathbf{p},E_\mathbf{p}/\hbar)\propto-n^{1/4}$. 
For the first time, an analytic result for the on-shell single-particle self-energy was found 
that contains the correct low-density limit,
i.e. a vanishing self-energy at $n=0$. This is a major progress compared to the well-known quasi-particle approximation that yields a finite damping width
even at zero density. The new on-shell single-particle damping width is
$-\mathrm{Im}\,\Sigma(\mathbf{p},E_\mathbf{p}/\hbar)=\left(3\,\Gamma^3\right)^{1/4}\,k_\mathrm{B}T$. Since it is derived in
Born approximation, i.e. no collective excitations contribute to the damping mechanisms, this quantity is called the non-collective damping width.
By comparison of the numerical results to the new analytic formula, 
the parameter $\eta=\sqrt{\kappa a_\mathrm{B}}$ was identified to separate the regime of
non-collective damping ($\eta\ll 1$) from the regime, where the coupling between single particle states and collective excitations 
dominantly determine the single particle damping ($\eta\gg 1$) at small momenta. This analysis complements earlier work on the electron spectral
function based on the plasmon-pole approximation.

For $\eta\ll 1$, the analytic formula~(\ref{eqn:analyticSigma_p-w})
is a good approximation for the self-energy. 
Furthermore, the use of the analytic formula for the self-energy as an initialization of
the iterative algorithm leads to significantly faster convergence as compared to other methods, where a Gaussian ansatz is used as the
initial spectral function.

The non-collective damping is a purely classical result, no powers of $\hbar$ appear.
This is fundamentally different from the quasi-particle approximation to the imaginary part of the self-energy
which has no classical limit, i.e. the self-energy diverges in the limit $\hbar\to 0$. It could be shown that this problem,
as well as the paradox of being density independent, stem from the inherently 
inconsistent treatment of the self-energy in the quasi-particle approximation. The long-time open question of the classical limit of the
single-particle self-energy can now be regarded as settled.

The results reported in this work are of paramount importance 
for many-particle theory and applications to dense plasmas.
In particular, simple analytic expressions for
the single-particle spectral function and self-energy in the classical and in the degenerate limit are needed to construct Pad\'e-like
interpolation formulae that cover the complete density-temperature plane. Such expressions would greatly simplify the calculation of equation of
state, transport and optical properties of dense, high energy plasmas, solid state devices but also nuclear, hadronic,
and partonic matter, and provide benchmarks for
numerical approaches, i.e. simulation techniques.
One part of this task, the analytic formula for non-degenerate dilute plasmas has been accomplished in this work.

\ack
  This work was performed with financial support from the German Research Society (DFG) under grant SFB 652 (Collaborative Research Center ``Strong Correlations and Collective Phenomena in Radiation Fields: Coulomb
  Systems, Clusters, and Particles''). Special thanks go to G. R\"opke and A. Wierling for many suggestions and comments on the manuscript
  and to W.-D. Kraeft for stimulating discussions.

\appendix
\section{Details on the $GW^{(0)}$-approximation\label{app:GW}}
Throughout the appendix, the Rydberg system of units will be applied to keep the formulae short and readable. In these units $\hbar=k_\mathrm{B}=1$, $e^2=2$, $\epsilon_0=1/4\pi$, and $m=1/2$.

We start from the representation of the self-energy in terms of the full Green function $G(\mathbf{p},z_\nu)$, the dynamically screened potential 
$W(\mathbf{q},\omega_\mu)$ and the vertex function $\Gamma(\mathbf{p},\mathbf{p}+\mathbf{q};z_\nu,z_\nu+\omega_\mu)$, given by the diagram
\begin{equation}
  \Sigma(\mathbf{p},z_\nu)=\parbox{80pt}{%
  \includegraphics[width=80pt,angle=0,clip]{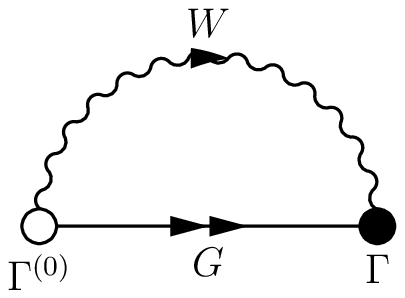}~.%
  \\}
  \label{appeqn:Sigma_GWGamma}
\end{equation}
In the $GW$-approximation, the vertex is replaced by the bare vertex $\Gamma^{(0)}=e$, i.e. the charge of the considered particles, electrons in
this case,
\begin{eqnarray}
  \Sigma(\mathbf{p},z_\nu)&= \parbox{80pt}{%
  \includegraphics[width=80pt,angle=0,clip]{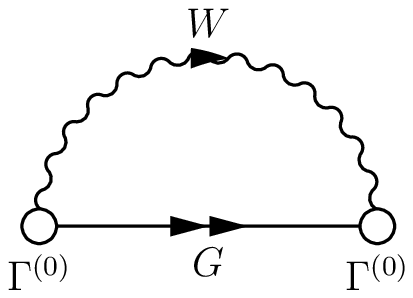}%
  \\}\\
  &= -T\sum_{\mathbf{q},\omega_\mu} G(\mathbf{p}-\mathbf{q},z_\nu-\omega_\mu)\,W(\mathbf{q},\omega_\mu)~,
  \label{appeqn:Sigma_GW}
\end{eqnarray}
which is equation (\ref{eqn:sigmaGW0_Matsubara}).
The dynamically screened interaction is taken in the random phase approximation \cite{AristaBrandt:PRA29_1984},
\begin{eqnarray}
  W^{(0)}(\mathbf{q},\omega_\mu)&= \frac{V(q)}{\epsilon_\mathrm{RPA}(\mathbf{q},\omega_\mu)}~,\\
  \epsilon_\mathrm{RPA}(\mathbf{q},\omega+\mathrm{i}\delta)&=1-V(q)\,\Pi_\mathrm{RPA}(\mathbf{q},\omega+\mathrm{i}\delta)\\
  \Pi_\mathrm{RPA}(\mathbf{q},\omega+\mathrm{i}\delta)&=-\sum_\mathbf{k}\frac{n_\mathrm{F}(\varepsilon_{\mathbf{k}+\mathbf{q}/2})-n_\mathrm{F}(\varepsilon_{\mathbf{k}-\mathbf{q}/2})}{\omega+\mathrm{i}\delta+\varepsilon_{\mathbf{k}-\mathbf{q}/2}-\varepsilon_{\mathbf{k}+\mathbf{q}/2}}~.
  \label{appeqn:epsPiRPA}
\end{eqnarray}
Using the spectral representations of both the Green function (\ref{eqn:spectral_representation_G}) and the screened interaction in RPA,
\begin{equation}
  W^{(0)}(\mathbf{q},z)=V(q)\left(
  1+\int_{-\infty}^{\infty}\frac{\mathrm{d}\omega}{\pi}\frac{\mathrm{Im}\,\epsilon^{-1}_\mathrm{RPA}(\mathbf{q},\omega+i\delta)}{z-\omega}
  \right)~,
  \label{eqn:spectral_representation_W0}
\end{equation}
leads to
\begin{eqnarray}
  \nonumber
  \Sigma(\mathbf{p},z_\nu)&= 
  -T\sum_{\mathbf{q},\omega_\mu}
  V(q)
  \int_{-\infty}^{\infty}\frac{\mathrm{d}\omega}{2\pi}
  \frac{A(\mathbf{p}-\mathbf{q},\omega)}{z_\nu-\omega_\mu-\omega}\\
  &\qquad\times\left(1+
  \int_{-\infty}^{\infty}\frac{\mathrm{d}\omega'}{\pi}
  \frac{\mathrm{Im}\,\epsilon^{-1}_\mathrm{RPA}(\mathbf{q},\omega')}{\omega_\mu-\omega'}\right)~,
  \label{eqn:sigmaGW0_spectral_representation}
\end{eqnarray}
and after summation of the Bosonic Matsubara frequencies,
\begin{eqnarray}
  \fl
  \nonumber
  \Sigma(\mathbf{p},z_\nu)=
  \sum_{\mathbf{q}}
  V(q)
  \int_{-\infty}^{\infty}\frac{\mathrm{d}\omega''}{2\pi}
  A(\mathbf{p}-\mathbf{q},\omega'')
  \\
  \nonumber
  \times\bigg(
  1-n_\mathrm{F}(\omega'')+
  \int_{-\infty}^{\infty}\frac{\mathrm{d}\omega'}{\pi}\frac{\mathrm{Im}\,\epsilon^{-1}_\mathrm{RPA}(\mathbf{q},\omega')\left[
  n_\mathrm{B}(\omega')+1-n_\mathrm{F}(\omega'')
   \right]}{z_\nu-\omega'-\omega''} \bigg)~,\\
  \label{eqn:sigmaGW0_eval010}
\end{eqnarray}
is obtained. This expression contains the Hartree-Fock self-energy of the interacting system,
\begin{equation}
  \Sigma^\mathrm{HF}_\mathrm{int}(\mathbf{p})=-\sum_{\mathbf{q}}
  \int_{-\infty}^{\infty}\frac{\mathrm{d}\omega}{2\pi} A(\mathbf{p}-\mathbf{q},\omega) n_\mathrm{F}(\omega)
  V(q)~,
  \label{eqn:sigmaFock_interacting}
\end{equation}
and the \textit{correlated} self-energy
\begin{eqnarray}
  \fl
  \nonumber
  \Sigma^\mathrm{corr}(\mathbf{p},z_\nu)=
  \sum_{\mathbf{q}}
  V(q)
  \int_{-\infty}^{\infty}\frac{\mathrm{d}\omega''}{2\pi}
  A(\mathbf{p}-\mathbf{q},\omega'')
  \\
  \times
  \int_{-\infty}^{\infty}\frac{\mathrm{d}\omega'}{\pi}\frac{\mathrm{Im}\,\epsilon^{-1}_\mathrm{RPA}(\mathbf{q},\omega')\left[
  n_\mathrm{B}(\omega')+1-n_\mathrm{F}(\omega'')
   \right]}{z_\nu-\omega'-\omega''}~.
  \label{eqn:sigmaGW0_eval015}
\end{eqnarray}
For convenience, we skip the upper index ``corr'' in the following and only distinguish between the frequency dependent self-energy
$\Sigma(\mathbf{p},\omega+i\delta)$ and the Hartree-Fock term $\Sigma^\mathrm{HF}_\mathrm{int}(\mathbf{p})$, in the following.

After analytic continuation $z_\nu\to z=\omega+\mathrm{i}\delta, \delta\to 0$, the imaginary part of the correlated self-energy is evaluated using Dirac's identity $\lim_{\delta\to0}1/(x\pm \mathrm{i}\delta)=\mathcal P1/x\mp
\mathrm{i}\pi\,\delta(x)$,
\begin{eqnarray}
  \fl\nonumber
  \mathrm{Im}\,\Sigma(\mathbf{p},\omega+\mathrm{i}0^+)= 
  \frac{1}{n_\mathrm{F}(\omega)}\sum_{\mathbf{q}}\int_{-\infty}^{\infty}\frac{\mathrm{d}\omega'}{2\pi}V_\mathrm{ee}(q)
  A(\mathbf{p}-\mathbf{q},\omega-\omega')\,
  \\
  \qquad\times
  \mathrm{Im}\,\epsilon^{-1}_\mathrm{RPA}(\mathbf{q},\omega')\,
  n_\mathrm{B}(\omega')\,n_\mathrm{F}(\omega-\omega')~,
  \label{eqn:sigmacorr_010app}
\end{eqnarray}
where the exact relation 
$n_\mathrm{B}(\omega')+1-n_\mathrm{F}(\omega-\omega')=-n_\mathrm{B}(\omega')\,n_\mathrm{F}(\omega-\omega')/n_\mathrm{F}(\omega)$ was used.
This equation is given as (\ref{eqn:sigmacorr_010}) in the main text.

\section{Analytic self-energy for the classical one-component plasma\label{app:derivation_analyticSigma}}


In the high temperature limit $k_\mathrm{B}T\gg E_\mathrm{F}$, we 
replace the Fermi Dirac distributions in the self-energy equation (\ref{eqn:sigmacorr_010}) as well as in the dielectric function
(\ref{eqn:epsPiRPA}) by the
Maxwell-Boltzmann distribution, $n_\mathrm{F}(\varepsilon_{\mathbf{k}})\to f(k)=\frac{n\Lambda^3}{2}\exp(-\varepsilon_{\mathbf{k}}/T)$ 
with the thermal de-Broglie wavelength $\Lambda=(4\pi/T)^{1/2}$. 
In this approximation, the polarization function takes the form \cite{FennelWilfert_AnnPhysL32_265_1974},
\begin{eqnarray}
  \fl
  \nonumber
    \mathrm{Re}\,\Pi_\mathrm{RPA}(\mathbf{q},\omega)=\frac{n}{2\,q\,T}\bigg[ \left( \frac{\omega}{q}-q \right)
    \,_1F_1\left( 1,3/2, -\left( \frac{\omega}{2q\sqrt{T}}-
    \frac{q}{2\sqrt{T}} \right)^2 \right)-\\
    \quad\qquad\left( \frac{\omega}{q}+q \right)
    \,_1F_1\left( 1,3/2, -\left( \frac{\omega}{2q\sqrt{T}}+\frac{q}{2\sqrt{T}} \right)^2 \right)\bigg]\\
    \fl
    \nonumber
    \mathrm{Im}\,\Pi_\mathrm{RPA}(\mathbf{q},\omega)=
    \frac{T\,n\Lambda^3}{8\pi\,q}\bigg[ \exp\left(-\left(
    \frac{\omega}{2q\sqrt{T}}+\frac{q}{2\sqrt{T}} \right)^2\right)-\\
    \quad\quad\quad\quad
    \exp\left(-\left( \frac{\omega}{2q\sqrt{T}}-\frac{q}{2\sqrt{T}} \right)^2 \right)\bigg]\\
    =-\frac{T\,n\Lambda^3}{8\pi\,q}\exp\left( -\frac{\omega}{2T} \right)\,\exp\left( -\left(
    \frac{\omega^2}{4q^2T}+\frac{q^2}{4T} \right) \right)\frac{1}{n_\mathrm{B}(\omega)}~.
  \label{eqn:PiRPA_classical}
\end{eqnarray}
Then, the imaginary part of the self-energy writes
\begin{eqnarray}
  \fl\nonumber
  \mathrm{Im}\,\Sigma(\mathbf{p},\omega+\mathrm{i}\delta)= 
  \frac{2\,\kappa^2}{\pi^{3/2}}\sqrt{T}\,\int_{-1}^{1}\!\mathrm{d}\cos\theta\,\int_{0}^{\infty}\!\mathrm{d}q\,\int_{-\infty}^{\infty}\!\mathrm{d}\omega'\\
  \nonumber
  \times\frac{1}{q^3}
  \frac{\exp\left( -\left( \frac{\omega'^2}{4q^2T}+\frac{q^2}{4T} \right) \right)}{|\epsilon(q,\omega')|^2}\,
  \exp\left( \frac{\omega'}{2T} \right)
  \\
  \nonumber
  \times\frac{\mathrm{Im}\,\Sigma(\mathbf{p}-\mathbf{q},\omega+\mathrm{i}\delta-\omega')}{\left[\omega-\omega'-\varepsilon_{\mathbf{p}-\mathbf{q}}-\mathrm{Re}\,\Sigma(\mathbf{p}-\mathbf{q},\omega-\omega')\right]^2+\left[\mathrm{Im}\,\Sigma(\mathbf{p}-\mathbf{q},\omega-\omega')\right]^2}.\\
\end{eqnarray}
Furthermore, the Born approximation is applied, i.e. the dielectric function in the denominator is replaced by the Debye expression,
$\epsilon_\mathrm{D}(q,0)=1+\kappa^2/q^2$. 
Diagrammatically, the self-energy in this approximation is written as
\begin{equation}
  \Sigma(\mathbf{p},z_\nu)=\parbox{80pt}{%
  \includegraphics[width=80pt,angle=0,clip]{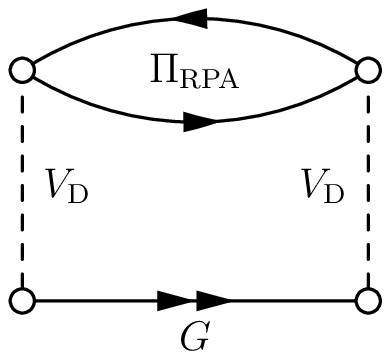}
  \\
  }~.
  \label{appeqn:sigma_born_rainbow}
\end{equation}
$V_\mathrm{D}$ denotes the Debye potential $V_\mathrm{D}(q)=e^2/\epsilon_0(q^2+\kappa^2)\Omega_0$.

Since the main contribution to the $q-$integral stems from momenta $q<\kappa$, we neglect the transfer wavenumber $q$ in the argument of the
self-energy on the r.h.s. and write
\begin{eqnarray}
  \fl
  \nonumber
  \mathrm{Im}\,\Sigma(\mathbf{p},\omega+\mathrm{i}\delta)=
  \frac{2\,\kappa^2}{\pi^{3/2}}\sqrt{T}\,\int_{-1}^{1}\!\mathrm{d}\cos\theta\,\int_{0}^{\infty}\!\mathrm{d}q\,\int_{-\infty}^{\infty}\!\mathrm{d}\omega'\\
  \nonumber
  \times\frac{1}{q^3}
  \frac{\exp\left( -\left( \frac{\omega'^2}{4q^2T}+\frac{q^2}{4T} \right) \right)}{[1+\frac{\kappa^2}{q^2}]^2}\,\exp\left(
  \frac{\omega'}{2T} \right)\\
  \times
  \frac{\mathrm{Im}\,\Sigma(\mathbf{p},\omega+\mathrm{i}\delta-\omega')}{\left[\omega-\omega'-\varepsilon_{\mathbf{p}-\mathbf{q}}-\mathrm{Re}\,\Sigma(\mathbf{p},\omega-\omega')\right]^2+\left[\mathrm{Im}\,\Sigma(\mathbf{p},\omega-\omega')\right]^2}
  \label{eqn:0100}
\end{eqnarray}
Furthermore, we neglect the term $q^2/4T$ in the exponential which is small for high temperatures and for $q<\kappa$,

\begin{eqnarray}
  \fl
  \nonumber
  \mathrm{Im}\,\Sigma(\mathbf{p},\omega+\mathrm{i}\delta)=
  \frac{2\,\kappa^2}{\pi^{3/2}}\sqrt{T}\,\int_{-1}^{1}\!\mathrm{d}\cos\theta\,\int_{0}^{\infty}\!\mathrm{d}q\,\int_{-\infty}^{\infty}\!\mathrm{d}\omega'\\
  \nonumber
  \times
  \frac{q}{[q^2+\kappa^2]^2}\,\exp\left( -\frac{\omega'^2}{4q^2T}\right)\exp\left(
  \frac{\omega'}{2T} \right)\\
  \times
  \frac{\mathrm{Im}\,\Sigma(\mathbf{p},\omega+\mathrm{i}\delta-\omega')}{\left[\omega-\omega'-\varepsilon_{\mathbf{p}-\mathbf{q}}-\mathrm{Re}\,\Sigma(\mathbf{p},\omega-\omega')\right]^2+\left[\mathrm{Im}\,\Sigma(\mathbf{p},\omega-\omega')\right]^2}
  \label{eqn:0150}
\end{eqnarray}
This is equation was given in section~\ref{sec:analytic} as (\ref{eqn:ImSBorn}).

Now, the integration over the angle $\theta$ can be performed as
\begin{eqnarray}
  \fl
  \nonumber
   \mathrm{Im}\,\Sigma(p,\omega+\mathrm{i}\delta)=
   \frac{2\,\kappa^2}{\pi^{3/2}}\sqrt{T}\,\int_{-1}^{1}\!\mathrm{d}\cos\theta\,\int_{0}^{\infty}\!\mathrm{d}q\,\int_{-\infty}^{\infty}\!\mathrm{d}\omega'
  \frac{q}{[q^2+\kappa^2]^2}\\
  \times\exp\left( -\frac{\omega'^2}{4q^2T}\right)\exp\left(
  \frac{\omega'}{2T} \right)
  \frac{\mathrm{Im}\,\Sigma(\mathbf{p},\omega+\mathrm{i}\delta-\omega')}{4p^2q^2}
  \\
  \fl
  \nonumber
  \times\bigg[\left(\frac{\omega-\omega'-p^2-q^2+\mu-\mathrm{Re}\,\Sigma(p,\omega-\omega')}{2pq}+\,\cos\theta\right)^2+\left(
  \frac{\mathrm{Im}\,\Sigma(\mathbf{p},\omega+\mathrm{i}\delta-\omega')}{2pq}\right)^2\bigg]^{-1}\\
  \fl
  \nonumber
  =\frac{\kappa^2}{\pi^{3/2}\,p}\sqrt{T}\,\int_{0}^{\infty}\!\mathrm{d}q\,\int_{-\infty}^{\infty}\!\mathrm{d}\omega'\,
  \frac{1}{[q^2+\kappa^2]^2}\,\exp\left( -\frac{\omega'^2}{4q^2T}\right)\exp\left(
  \frac{\omega'}{2T} \right)
  \\
  \nonumber
  \quad\times\bigg[ 
  \arctan\left(
  \frac{(p+q)^2-\omega-\mu+\mathrm{Re}\,\Sigma(\mathbf{p},\omega-\omega')}{\mathrm{Im}\,\Sigma(\mathbf{p},\omega+\mathrm{i}\delta-\omega')}\right)\\
  \qquad
  \arctan\left(\frac{(p-q)^2-\omega-\mu+\mathrm{Re}\,\Sigma(\mathbf{p},\omega-\omega')}{\mathrm{Im}\,\Sigma(\mathbf{p},\omega+\mathrm{i}\delta-\omega')}
  \right)
  \bigg]~,
\end{eqnarray}
where the integral $\int dx/\left[(a+x)^2+b^2\right]=b^{-1}\arctan\left( \left( a+x \right)/b \right)$ was used.

In the limit $q\to 0$ the identity
\begin{equation}
  \lim_{q\to 0}\frac{1}{2q\sqrt{T}}\,\mathrm{e}^{-\omega'^2/4q^2T}=\sqrt{\pi}\,\delta(\omega')~,
  \label{eqn:exp-delta}
\end{equation}
allows us to perform the frequency integration,
\begin{eqnarray}
  \fl
  \nonumber
 \mathrm{Im}\,\Sigma(p,\omega+\mathrm{i}\delta)=
 \frac{\kappa^2}{\pi^{3/2}\,p}\sqrt{T}\,\int_{0}^{\infty}\!\mathrm{d}q\,
 \frac{2q\sqrt{T}}{[q^2+\kappa^2]^2}
 \int_{-\infty}^{\infty}\!\mathrm{d}\omega'\,\exp\left( \frac{\omega'}{2T} \right)
  \,\frac{\exp\left( -\frac{\omega'^2}{4q^2T}\right)}{2q\sqrt{T}}\\
  \nonumber
  \fl
  \quad\times\bigg[ 
  \arctan\left(
  \frac{(p+q)^2-\omega-\mu+\mathrm{Re}\,\Sigma(\mathbf{p},\omega-\omega')}{\mathrm{Im}\,\Sigma(\mathbf{p},\omega+\mathrm{i}\delta-\omega')}\right)\\
  \nonumber
  -
  \arctan\left(\frac{(p-q)^2-\omega-\mu+\mathrm{Re}\,\Sigma(\mathbf{p},\omega-\omega')}{\mathrm{Im}\,\Sigma(\mathbf{p},\omega+\mathrm{i}\delta-\omega')}
  \right)
  \bigg]\\
  \nonumber
  \fl
  =
  \frac{2\,\kappa^2\,T}{\pi^{3/2}\,p\,}\,\int_{0}^{\infty}\!\mathrm{d}q\,
 \frac{q}{[q^2+\kappa^2]^2}
 \int_{-\infty}^{\infty}\!\mathrm{d}\omega'\,\exp\left( \frac{\omega'}{2T} \right)
  \,\sqrt{\pi}\,\delta(\omega')\\
  \nonumber
  \fl
  \quad\times\bigg[ 
  \arctan\left(
  \frac{(p+q)^2-\omega-\mu+\mathrm{Re}\,\Sigma(\mathbf{p},\omega-\omega')}{\mathrm{Im}\,\Sigma(\mathbf{p},\omega+\mathrm{i}\delta-\omega')}\right)\\
  \nonumber
  - \arctan\left(\frac{(p-q)^2-\omega-\mu+\mathrm{Re}\,\Sigma(\mathbf{p},\omega-\omega')}{\mathrm{Im}\,\Sigma(\mathbf{p},\omega+\mathrm{i}\delta-\omega')}
  \right)
  \bigg]\\
  \nonumber
  \fl
  =
  \frac{2\,\kappa^2\,T}{\pi\,p\,}\,\int_{0}^{\infty}\!\mathrm{d}q\,
  \frac{q}{[q^2+\kappa^2]^2}\\
  \nonumber
  \fl
  \quad\times\bigg[ 
  \arctan\left(
  \frac{(p+q)^2-\omega-\mu+\mathrm{Re}\,\Sigma(\mathbf{p},\omega)}{\mathrm{Im}\,\Sigma(\mathbf{p},\omega+\mathrm{i}\delta)}\right)\\
  -\arctan\left(\frac{(p-q)^2-\omega-\mu+\mathrm{Re}\,\Sigma(\mathbf{p},\omega)}{\mathrm{Im}\,\Sigma(\mathbf{p},\omega+\mathrm{i}\delta)}
  \right)
  \bigg]
\end{eqnarray}
Using the following power expansion of the $\arctan$-function
\begin{equation}
  \arctan(1+x)=\frac{\pi}{4}+\frac{x}{2}+\mathcal{O}(x^3)~,
\end{equation}
i.e.
\begin{eqnarray}
  \nonumber
  \fl
\arctan\left(
  \frac{(p+q)^2-\omega-\mu+\mathrm{Re}\,\Sigma(\mathbf{p},\omega)}{\mathrm{Im}\,\Sigma(\mathbf{p},\omega+\mathrm{i}\delta)}\right)-\\
  \nonumber
  \arctan\left(\frac{(p-q)^2-\omega-\mu+\mathrm{Re}\,\Sigma(\mathbf{p},\omega)}{\mathrm{Im}\,\Sigma(\mathbf{p},\omega+\mathrm{i}\delta)}
  \right)
  \\
  \fl
  = 4pq\left\{\mathrm{Im}\,\Sigma(\mathbf{p},\omega+\mathrm{i}\delta)\left[1+\left(\frac{p^2-\omega-\mu+\mathrm{Re}\,\Sigma(\mathbf{p},\omega)}
  {\mathrm{Im}\,\Sigma(\mathbf{p},\omega+\mathrm{i}\delta)}\right)^2\right]\right\}^{-1}+\mathcal{O}(q^3)~,
\end{eqnarray}
we obtain
\begin{eqnarray}
  \nonumber
  \fl
  \mathrm{Im}\,\Sigma(p,\omega+\mathrm{i}\delta)=
  \frac{2\,\kappa^2\,T}{\pi\,p}\,\int_{0}^{\infty}\!\mathrm{d}q\,
  \frac{q}{[q^2+\kappa^2]^2}\\
  \times
  4pq\left\{\mathrm{Im}\,\Sigma(\mathbf{p},\omega+\mathrm{i}\delta)\left[1+\left(\frac{p^2-\omega-\mu+\mathrm{Re}\,\Sigma(\mathbf{p},\omega)}
  {\mathrm{Im}\,\Sigma(\mathbf{p},\omega+\mathrm{i}\delta)}\right)^2\right]\right\}^{-1}<++>,\\
\end{eqnarray}
which can be turned into
\begin{eqnarray}
  \nonumber
  \fl
  \left[ \mathrm{Im}\,\Sigma(\mathbf{p},\omega+\mathrm{i}\delta) \right]^2+\left[ p^2-\omega-\mu+\mathrm{Re}\,\Sigma(\mathbf{p},\omega) \right]^2
  &=
  \frac{8\,\kappa^2\,T}{\pi}\int_0^{\infty}\!\mathrm{d}q\,\frac{q^2}{(q^2+\kappa^2)^2}
  \\&
  =2\kappa\,T~.
\end{eqnarray}
To solve this single equation for the two unknown $\mathrm{Re}\,\Sigma(\mathbf{p},\omega)$ and
$\mathrm{Im}\,\Sigma(\mathbf{p},\omega+\mathrm{i}\delta)$, we make use of the spectral representation of the Green function
\begin{eqnarray}
  \nonumber
  \fl
  G(\mathbf{p},z)=\left[ z-p^2-\Sigma(\mathbf{p},z) \right]^{-1}
  =\int_{-\infty}^{\infty}\!\frac{\mathrm{d}\omega}{2\pi}\frac{A(\mathbf{p},\omega)}{z-\omega}\\
  \fl
  =\int_{-\infty}^{\infty}\!\frac{\mathrm{d}\omega}{2\pi}\frac{\mathrm{Im}\,\Sigma(\mathbf{p},\omega+\mathrm{i}\delta)}{z-\omega}
  \bigg[
  \left( \mathrm{Im}\,\Sigma(\mathbf{p},\omega+\mathrm{i}\delta) \right)^2
  +\left(\omega+\mu-p^2-\mathrm{Re}\,\Sigma(\mathbf{p},\omega) \right)^2
  \bigg]^{-1}\\
  \fl
  =
  \int_{-\infty}^{\infty}\!\frac{\mathrm{d}\omega}{2\pi}\frac{\mathrm{Im}\,\Sigma(\mathbf{p},\omega+\mathrm{i}\delta)}{z-\omega}\frac{1}{2\kappa\,T}
  =
  \frac{\Sigma(\mathbf{p},z)}{2\kappa T}~.
\end{eqnarray}
In the last step we also used the spectral representation of the correlated self-energy.
The last equation has the solution
\begin{equation}
  \Sigma(\mathbf{p},z)=\frac{z-p^2+\mu}{2}\pm\left[ \left( \frac{z-p^2+\mu}{2} \right)^2-2\kappa\,T\right]^{1/2}~.
  \label{eqn:0200}
\end{equation}
With $z=\omega+\mathrm{i}\delta$ and having in mind that $\mathrm{Im}\,\Sigma(\mathbf{p},\omega+\mathrm{i}\delta)<0$ for $\delta>0$, we finally find
\begin{equation}
  \fl
  \Sigma(\mathbf{p},\omega+\mathrm{i}\delta)=\frac{\omega+\mu-p^2}{2}-\mathrm{sign}(\omega+\mu-p^2)\left[\left( \frac{\omega+\mu+\mathrm{i}\delta-p^2}{2} \right)^2-
  2\kappa\,T \right]^{1/2}\!\!,
  \label{eqn:analyticSigma_p-w_app}
\end{equation}
i.e. equation (\ref{eqn:analyticSigma_p-w}).

\section{Details for the quasi-particle self-energy\label{app:p0w0}}
We start from (\ref{eqn:ImSBorn}) for the imaginary part of the self-energy using the Born approximation for the screened interaction
potential:
\begin{eqnarray}
  \fl
  \nonumber
  \mathrm{Im}\,\Sigma(\mathbf{p},\omega+\mathrm{i}\delta)= -\frac{\sqrt{T}\kappa^2}{\pi^{3/2}}\,\int_{-1}^{1}\!
  \mathrm{d}\cos\theta\,\int_{0}^{\infty}\!\mathrm{d}q\,
  \int_{-\infty}^{\infty}\!\mathrm{d}\omega'\,\frac{q}{\left( q^2+\kappa^2 \right)^2}
  \\
  \times
  A(\mathbf{p}-\mathbf{q},\omega-\omega')\,
  \exp\left( \frac{\omega'}{2T} \right)\,\exp\left( -\frac{\omega'^2}{4q^2T}\right)~.
  \label{appeqn:ImSBorn}
\end{eqnarray}

By assuming a frequency and momentum independent self-energy 
$\Sigma(\mathbf{p},\omega)\equiv \Sigma(0,E_0/\hbar)$,
this becomes
\begin{eqnarray}
  \nonumber
  \fl
 \mathrm{Im}\,\Sigma(0,E_0/\hbar)
 =\frac{2\sqrt{T}\kappa^2}{\pi^{3/2}}\,\int_{-1}^{1}\! \mathrm{d}\cos\theta\,\int_{0}^{\infty}\!\mathrm{d}q\,
 \int_{-\infty}^{\infty}\!\mathrm{d}\omega'\,\frac{q}{\left( q^2+\kappa^2 \right)^2}\\
  \times\frac{\mathrm{Im}\,\Sigma(0,E_0/\hbar)}{\left[
  \omega'+q^2\right]^2+\left[
  \mathrm{Im}\,\Sigma(0,E_0/\hbar) \right]^2}
  \exp\left( \frac{\omega'}{2T} \right)\,\exp\left( -\frac{\omega'^2}{4q^2T}\right)~.
  \label{eqn:app_scalinglawderiv_010}
\end{eqnarray}
Since the self-energy is assumed to be independent of the frequency, 
the real part of the correlated self-energy vanishes exactly. For the
Hartree-Fock part of the self-energy is proportional to
$n\Lambda^3$ in the classical limit \cite{krae}, we also 
neglect this term,
since it gives contributions of higher order in $n$, whereas
we are only interested in the lowest order.

After eliminating $\mathrm{Im}\,\Sigma(0,E_0/\hbar)$ on both sides, performing the trivial integration over the angle $\theta$, which yields a
factor $2$, the frequency integration is performed by the help of \cite{abra}
\begin{equation}
  \int_0^{\infty}\frac{\exp(-t^2)\,\mathrm{d}t}{z-t}=\frac{\pi}{2\mathrm{i} z}\,\exp(-z^2)\,\mathrm{erfc}(-\mathrm{i}z)~,
  \label{appeqn:integral}
\end{equation}
leading to
\begin{eqnarray}
  \fl
  1=4\sqrt{\frac{T}{\pi^3}}\kappa^2\int_{0}^{\infty}
  \frac{\mathrm{d}q\, q}{(q^2+\kappa^2)^2}\int_{-\infty}^{\infty}\frac{\exp(-\omega'^2/4q^2T)\,\mathrm{d}\omega'}{[q^2+\omega']^2+[\mathrm{Im}\,\Sigma(0,E_0/\hbar)]^2}\\
  =-4\sqrt{\frac{T}{\pi^3}}\kappa^2\int_{0}^{\infty}
  \frac{\mathrm{d}q\,q}{(q^2+\kappa^2)^2}\frac{\pi\,\mathrm{Re}\,\left[
  \exp(-z^2)\,\mathrm{erfc}(\mathrm{i}z) \right]}{\mathrm{Im}\,\Sigma(0,E_0/\hbar)},
\end{eqnarray}
which is rewritten as
\begin{equation}
  \fl
  \mathrm{Im}\,\Sigma(0,E_0/\hbar)
  =-4\sqrt{\frac{T}{\pi}}\kappa^2\int_{0}^{\infty}
  \frac{\mathrm{d}q\,q}{(q^2+\kappa^2)^2}\,\mathrm{Re}\,\left[ \exp(-z^2)\,\mathrm{erfc}(\mathrm{i}z) \right],
  \label{eqn:app_scalinglawderiv_031}
\end{equation}
with
\begin{equation}
  z=\frac{q^2+\mathrm{i}\,\mathrm{Im}\,\Sigma(0,E_0/\hbar)}{2q\sqrt{T}}~.
  \label{eqn:app_scalinglawderiv_032}
\end{equation}
Equation (\ref{eqn:app_scalinglawderiv_031}) is given as (\ref{eqn:app_scalinglawderiv_030}) in section~\ref{sec:p0w0}.
It should be noted at this point that the integral converges only for 
finite $\kappa$, i.e. the Coulomb limit $\kappa\to 0$ does not yield a
finite result.

Most contributions to the integral stem from small values of the
wavevector, $q\lesssim \kappa$. Therefore, we may neglect the
real part of $z$ and write
$z=\mathrm{i}\,\mathrm{Im}\,\Sigma(0,E_0/\hbar)/2q\sqrt{T}$. Using the expansion 
\begin{equation}
  \lim_{x\to+\infty}\,\exp(x^2)\,\mathrm{erfc}(x)=\frac{1}{\sqrt{\pi}\,x}-\frac{1}{2\sqrt{\pi}x^2}+\mathcal{O}(x^{-3})
  \label{eqn:expansion_experfc_app}
\end{equation}
in lowest order only, the $q$-integral can be performed, resulting in
\begin{eqnarray}
  \fl
  \mathrm{Im}\,\Sigma(0,E_0/\hbar)=-4\sqrt{\frac{T}{\pi}}\kappa^2\int_{0}^{\infty}
  \frac{\mathrm{d}q\,q}{(q^2+\kappa^2)^2}\\
  \nonumber
  \times\mathrm{Re}\,\left[\exp(-(i\mathrm{Im}\,\Sigma(0,E_0/\hbar)/2q\sqrt{T})^2)\,\mathrm{erfc}(\mathrm{i}\cdot
  \mathrm{i}\mathrm{Im}\,\Sigma(0,E_0/\hbar)/2q\sqrt{T})\right]\\
  =-4\sqrt{\frac{T}{\pi}}\kappa^2\int_{0}^{\infty}\frac{\mathrm{d}q\, q}{(q^2+\kappa^2)^2}\\
  \nonumber
  \times\mathrm{Re}\,\left[\exp((-\mathrm{Im}\,\Sigma(0,E_0/\hbar)/2q\sqrt{T})^2)\,\mathrm{erfc}(-\mathrm{Im}\,\Sigma(0,E_0/\hbar)/2q\sqrt{T})\right]\\
  \simeq -2\sqrt{\frac{T}{\pi}}\kappa^2\int_{0}^{\infty}
  \frac{\mathrm{d}q\,q}{(q^2+\kappa^2)^2}\,\frac{1}{\sqrt{\pi}(-\mathrm{Im}\,\Sigma(0,E_0/\hbar)/2q\sqrt{T})}\\
  =\frac{8T\kappa^2}{\pi\mathrm{Im}\,\Sigma(0,E_0/\hbar)}\int_0^{\infty}\frac{\mathrm{d}q\,q^2}{(q^2+\kappa^2)^2}=
  \frac{8T\kappa^2}{\pi\mathrm{Im}\,\Sigma(0,E_0/\hbar)}\frac{\pi}{4 \kappa}~.
\end{eqnarray}
Finally, from the last line,
\begin{equation}
  \mathrm{Im}\,\Sigma(0,E_0/\hbar)=-\sqrt{2\,T\,\kappa}~,
\end{equation}
is obtained, and, after re-establishing SI units, equation (\ref{eqn:ImSQP}).


\section*{References}

\end{document}